\documentclass[journal]{IEEEtran}
\usepackage[flushleft]{threeparttable}
\usepackage{booktabs}
\ifCLASSINFOpdf
\else
\fi
\usepackage{cite}
\usepackage{amsmath,amssymb,amsfonts}
\usepackage{algorithmic}
\usepackage{graphicx}
\usepackage{textcomp}
\usepackage[flushleft]{threeparttable}
\usepackage{caption}
\usepackage{subcaption}
\usepackage{booktabs}
\usepackage{multirow}
\usepackage{hyperref}
\usepackage[font=small,skip=0pt]{subcaption}
\usepackage{tikz}
\usepackage[export]{adjustbox}

\begin{document}
\title{Consumer Grade Brain Sensing for \\Emotion Recognition}
\author{Payongkit Lakhan, Nannapas Banluesombatkul, Vongsagon Changniam, Ratwade Dhithijaiyratn,\\Pitshaporn Leelaarporn, Ekkarat Boonchieng, Supanida Hompoonsup and Theerawit Wilaiprasitporn
\thanks{This work was supported by Robotics AI and Intelligent Solution Project, PTT Public Company Limited, Thailand Research Fund and the Office of Higher Education Commission under Grant MRG6180028 and Junior Science Talent Project, NSTDA, Thailand.}
\thanks{P. Lakhan, N. Banluesombatkul, P. Leelaarporn and T. Wilaiprasitporn are members of the Bio-inspired Robotics and Neural Engineering Lab at the School of Information Science and Technology, Vidyasirimedhi Institute of Science \& Technology, Rayong, Thailand {\tt\small (corresponding author: theerawit.w at vistec.ac.th)}}
\thanks{V. Changniam is with the Department of Tool and Materials Engineering, King Mongkut's University of Technology Thonburi, Bangkok, Thailand}
\thanks{R. Dhithijaiyratn is with the Department of Electrical Engineering, Chulalongkorn University, Bangkok, Thailand}
\thanks{E. Boonchieng is with the Center of Excellence in Community Health Informatics, Chiang Mai University, Chiang Mai, Thailand}
\thanks{S. Hompoonsup is with the Learning Institute, King Mongkut's University of Technology Thonburi, Bangkok, Thailand}
}

%\markboth{Journal of \LaTeX\ Class Files,~Vol.~14, No.~8, August~2019}%
%{Shell \MakeLowercase{\textit{et al.}}: Bare Demo of IEEEtran.cls for IEEE Journals}

\maketitle

\begin{abstract}
For several decades, electroencephalography (EEG) has featured as one of the most commonly used tools in emotional state recognition via monitoring of distinctive brain activities. An array of datasets have been generated with the use of diverse emotion-eliciting stimuli and the resulting brainwave responses conventionally captured with high-end EEG devices. However, the applicability of these devices is to some extent limited by practical constraints and may prove difficult to be deployed in highly mobile context omnipresent in everyday happenings. In this study, we evaluate the potential of OpenBCI to bridge this gap by first comparing its performance to research grade EEG system, employing the same algorithms that were applied on benchmark datasets. Moreover, for the purpose of emotion classification, we propose a novel method to facilitate the selection of audio-visual stimuli of high/low valence and arousal. Our setup entailed recruiting 200 healthy volunteers of varying years of age to identify the top 60 affective video clips from a total of 120 candidates through standardized self assessment, genre tags, and unsupervised machine learning. Additional 43 participants were enrolled to watch the pre-selected clips during which emotional EEG brainwaves and peripheral physiological signals were collected. These recordings were analyzed and extracted features fed into a classification model to predict whether the elicited signals were associated with a high or low level of valence and arousal. As it turned out, our prediction accuracies were decidedly comparable to those of previous studies that utilized more costly EEG amplifiers for data acquisition.
\end{abstract}


\IEEEpeerreviewmaketitle

\section{Introduction}
\IEEEPARstart{E}{motion} plays an integral role in human social interactions and is typically evoked as a result of psychological or physiological responses to external stimuli. Due to the distinctly personal nature, emotions or affective states are traditionally assessed by psychologists through self report means and can be classified into well-defined categories \cite{russell1980}. Over the years, emotion studies have broadened from psycho-physiological focuses to engineering applications \cite{picard2000affective}, and numerous tools and algorithms have been developed in an attempt to tackle the challenges of emotion recognition encountered by the latter. In particular, the 21st century brought a dramatic increase in the number of investigative efforts into brain activities during emotional processing. This is in part owing to the advent of a non-invasive technique known as electroencephalography (EEG). A standard EEG device consists of multiple electrodes that can capture both spatial and temporal information, very much like a video capture \cite{reis2014methodological}. Other brain imaging techniques such as functional Magnetic Resonance Imaging (fMRI) might offer higher spatial resolution, but EEG devices come with a lower cost, higher temporal resolution, lighter weight, and simpler assembly. These attributes may enable researchers to monitor instantaneous changes in the brain as emotion is being elicited by extraneous influence.

Most emotion induction approaches make use of audio or visual stimuli to rouse emotions of varying arousal and valence levels on command. Apposite brainwave and physiological measures are gathered concurrently and these multimodal recordings are then subjected to analysis by an emotion recognition algorithm to assign the most likely emotion label, followed by an evaluation of overall prediction accuracy. This prototypical workflow is routinely adopted in emotion recognition research. One such study applied an EEG-based feature extraction technique to classify six basic emotions (anger, disgust, fear, happiness, sadness, and surprise) evoked during the viewing of pictures of facial expressions \cite{petrantonakis2009emotion}. In a similar fashion, Lin and colleagues \cite{lin2010eeg} extracted bio-signals using EEG and classified users’ affective states while listening to music into 4 groups namely, anger, pleasure, sadness, and joy. Nie and colleagues \cite{nie2011eeg} on the other hand utilized combined audio and visual stimuli in the form of 12 movie clips and classified the trajectory of emotions into positive and negative values using self-assessment manikin (SAM) and features extracted from EEG time waves. In lieu of external stimuli, Kothe and colleagues \cite{kothe2013} relied on the power of imagination to induce emotions. In their study, the participants were asked to recall memories that matched the description of targeted emotions, including positive and negative valence, during which their EEG signals were being recorded. In most cases, there is considerable room for improvement when it comes to prediction accuracy; these are equally true for studies that deploy either external or internal emotion-eliciting stimuli.

Several EEG-based benchmark databases for emotion recognition have been generated by independent groups; a fair number of which are freely available to the public. One of the first publicly accessible datasets, MAHNOB-HCI, was created to meet the growing demand from the scientific community for multifaceted information. The dataset contains EEG signals, physiological responses, facial expressions, and gaze data collected during participants' video viewing \cite{mahnob}. The affective clips were short section cuts from a collection of commercial films. The experimental design was built upon a preliminary study whereby online volunteers reported the emotions experienced during their watching of the clips. In the main study, the participants were instructed to submit a subjective rating on arousal, valence, dominance, and content predictability. Another dataset called Database for Emotion Analysis using Physiological Signals or DEAP was similarly motivated; it contains brainwave recordings from EEG in combination with peripheral physiological signals \cite{deap}. The stimulus of choice was an assortment of music videos that were pre-selected based on the demographic background of the participants, most of whom were European. These music videos were retrieved from a website that collects songs and music videos tagged with different emotion descriptors by users. Each of the music videos received a score from the participants on arousal, valence, dominance, liking, and familiarity.

As opposed to the previously mentioned datasets, a multimodal dataset built for the decoding of user physiological responses to affective multimedia content or DECAF compared the signals obtained from EEG to the recordings from ELEKTA Neuromag, a Magnetoencephalogram (MEG) sensor \cite{decaf}. Video selection was reliant on the affective notations reported by volunteers prior to the study. In the data collection phase, participants were told to rate the music videos and movie clips in terms of arousal, valence, and dominance. Other types of measurable signals that correlated with implicit affective states were simultaneously monitored and collected with horizontal Electrooculogram (hEOG), Electrocardiogram (ECG), Near-Infra-Red (NIR) facial videos, and trapezius-Electromyogram (tEMG). Constructed in 2015, SJTU Emotion EEG Dataset or SEED comprises EEG, ECG, EMG, EOC, and skin resistance (SC) data acquired from 15 Chinese students \cite{zheng2015investigating}. Fifteen emotion-charged clips were selected from a pool of popular locally produced Chinese films. The participants were directed to categorize the clips as either positive, neutral, or negative. However, the classification of emotions in this dataset did not include valence and arousal as labels. This study was an extension from the earlier study by the same research group where movie clips were being scored on three attributes i.e., valence, arousal, and dominance \cite{nie2011eeg}. The most recent dataset is possibly DREAMER, created using consumer grade EEG and ECG. The EEG system used was an Emotiv EPOC which is famous for its easy-to-wear design and wireless headset \cite{dreamer}. To gather the relevant data, the team instructed their recruits to fill in the SAM form to rate their levels of arousal, valence, and dominance. Short videos clips were chosen as the means of emotion elicitation and the emotions most likely to be invoked from these clips were predetermined. \newline 

\begin{table*}
\centering
\begin{threeparttable}
\caption{Comparison of consumer grade EEG devices on the market.} 
\begin{tabular}{l c c c c c c } 
\toprule
\toprule

Product Name & Price [USD] & Sampling Rate [Hz] & No. of Channels & Open Source & Raw Data & Scientific Validation\\
\midrule 

OpenBCI & 750/1800 & 250 &8/16 &Yes & Yes& MRCPs \cite{mdpi} \\
EMOTIV EPOC+ &	799	&256 &	14	& SDK & Yes*	&P300\cite{epoc_p300}, ERP\cite{audi_ERP}, emotion\cite{dreamer}\\
EMOTIV INSIGHT & 299 & 128 & 5 & SDK & Yes* & - \\
Myndplay Myndband & 200	& 512 &	3 & SDK*	&Yes & - \\
NeuroSky MindWave Mobile & 99 &512 & 1&
Yes&Yes&frontal EEG\cite{muse_neurosky}\\
InteraXon MUSE &	199 & 220 &	4 &	SDK &Yes& frontal EEG\cite{muse_neurosky}, ERP\cite{muse_ERP}  \\	
\bottomrule 
\bottomrule
\end{tabular}
\begin{tablenotes}
      \small \item \textit{*At additional cost, SDK stands for Software Development Kits}
    \end{tablenotes}
\label{lowcost}
\end{threeparttable}
\end{table*}

The key objective of our investigation was to evaluate the usability of an open-source consumer grade EEG amplifier, OpenBCI, in emotion recognition application. In particular, by adopting the same paradigm and classification algorithm as used by high-end EEG works, we were able to appraise the prediction accuracies of OpenBCI-derived data in a systematic manner against previous studies and found them to be more or less comparable \cite{deap,mahnob,dreamer}. We developed a classification model to predict whether the recorded EEG data had a high or low level of valence and arousal. The algorithm essentially imparts simple feature extraction (e.g., power spectral density (PSD)) and the classifier was constructed with a support vector machine (SVM) architecture. In sum, our classification results appeared on par if not better when compared to the results from benchmark datasets that were generated from high-end EEG devices.
Our two principal contributions can be summarized as follows:
\begin{itemize}
    \item We propose a robust stimulus selection method and validate a clustering approach for emotion labeling.
    \item We provide concrete evidence for the capability of consumer grade OpenBCI in emotion recognition study by quality assessing our performance accuracies against results from public repositories.
\end{itemize}
The remainder of this paper provides an overview of the rising presence of consumer grade EEG devices in emotion studies (Section II), methodology in Section III, results in Section IV, discussions in Section V, and conclusion in Section VI.

\section{Consumer Grade EEG for Emotion Recognition}
\textcolor{black}{According to our literature survey on recently published articles using \textit{``Low-Cost EEG Emotion"} as a search phrase, the majority of studies used EEG headsets from EMOTIV EPOC+ \cite{epoc}, while others used EMOTIV INSIGHT, MYNDPLAY \cite{myd}, NeuroSky \cite{nsk}, and MUSE \cite{muse}. Only a few studies on emotion recognition used OpenBCI  \cite{OpenBCI}. \autoref{lowcost} presents a comparison between a brand-new consumer grade EEG by the name of OpenBCI and other consumer grade EEGs on the market. Each product has been scientifically validated against high-end EEGs by measuring the established brain responses. Three studies validated EPOC+ against EEG from ANT Neuro \cite{ant}, Neuroscan\cite{neurosc}, and gtec\cite{gtec}. The studies included the measurement of P300, Event-Related Potential (ERP), and emotion, respectively. Frontal EEGs recorded from NeuroSky and MUSE were validated with two baseline EEGs named B-Alert  \cite{advanc} and Enobio \cite{neuroelec}. MUSE was also validated with the EEG from Brain Products GmbH \cite{brainproduct} for ERP research. Recently, one research group reported the performance of OpenBCI based on Texas Instrument ADS1299 (biopotential amplifier) using movement-related cortical potential measurement. In comparison to the product from Neuroscan\cite{neurosc}, there were no statistically significant differences between the two makes.}

\textcolor{black}{Research on emotion-related topics that adopt consumer grade EEG can be separated into two domains depending on the types of emotion-eliciting stimuli. The first domain consists of audio-visual stimulation (video clips). Two studies \cite{dreamer, amigos} set out to construct emotion datasets for public release and develop recognition algorithms for prospective applications. One research team integrated EMOTIV EPOC+ with a bracelet sensor \cite{e4} and eye tracker \cite{tobii} to produce a consumer grade emotion recognition system \cite{int}. An evolutionary computation was then proposed as a competitive algorithm in emotion recognition and EMOTIV INSIGHT was used for performance evaluation \cite{EC}. Furthermore, MUSE and certain other physiological datasets were used in another work involving boredom detection during video clip viewing \cite{kim}. Data collected from the MUSE headband were also shown to be able to predict the affective states and their magnitudes \cite{Becker2017}. As an example of its practical application, a real-time emotion recognition framework was recently proposed, using EMOTIV EPOC+ to record participants' brainwaves as they were being shown Chinese film excerpts \cite{real}. The second domain involves auditory stimulation, most often music, as the means to invoke desired emotions. The use of EMOTIV EPOC+ was demonstrated for the automatic detection of brain responses from three types of music (neutral, joyful, and melancholic). EEG connectivity features were reported to play an essential role in the recognition tasks \cite{shaba}. Another study, using the same device, observed a correlation of EEG with classical music rhythm \cite{rhythm,rhythm2}. Music therapy is one of the applications emanating from music-elicited emotion \cite{therapy}. Moreover, in the study of auditory-related emotion, one research group reported the development of envisioned speech recognition using EEG \cite{envisioned}.}

\textcolor{black}{Motivated by the findings of the aforementioned research domains on emotion-related works, researchers from other fields have also used consumer grade EEG devices. Examples of such studies are provided in this paragraph to enable the reader to get a sense of the impact of consumer grade devices. Two papers have reviewed past literature on the study of human behavior using brain signals for consumer neuroscience and neuromarketing applications \cite{applying, marketing}. In addition, a consumer-related research study showcased the effect of color priming on dress shopping, measured by EEG \cite{shopping}. Another group reported the feasibility of using asymmetry in frontal EEG band power (especially the Alpha band) as a metric for user engagement while reading the news \cite{engagement}. Band power asymmetry was also introduced as a study feature on the effects of meditation on emotional responses \cite{meditation}. Some groups explored simultaneous EEG recordings from multiple subjects such as the research on synchronized-EEGs while students were studying in the same classroom \cite{classroom} and similar research on the emotional engagement between a performer and multiple audiences \cite{performer}. Other research groups have studied longitudinal EEG data from individual subjects such as an investigation involving 103-day single channel EEG data on emotion recognition in daily life \cite{daily}. Nowadays, fundamental knowledge on emotion recognition or affective computing using brain wave activity has been applied in broad areas such as studies of people with depression \cite{depression}, stress in construction workers \cite{stress}, and interactions between ambient temperature in building and the occupant \cite{human}.}

\textcolor{black}{As previously mentioned, this paper describes a feasibility study of emotion recognition by OpenBCI. In recent years, there have been a few related works, one of which demonstrated the usefulness of OpenBCI as a biofeedback instrument (EEG, EMG) for an application titled IBPoet \cite{ibpoet}. Three parties participated in the IBPoet demonstration: the reader, the readee, and the audience. Once the reader had relayed the selected poem, the readee sent biofeedback (emotional responses) to the reader via a vibration band and heat glove. The reader and the system adapted according to the emotional responses to give the audience a greater feeling of pleasure. Similarly, another research group used the same device to evaluate and practice oral presentation \cite{oral}. The other findings are related to typical research on emotion recognition, but the numbers of experimental subjects were very small (some have several subjects) \cite{OpenBCIE, music2, stress2, dailyvalence}. Moreover, EEG software for automated emotion recognition to support consumer grade devices, including OpenBCI, was proposed in late 2018 \cite{software}. Thus, it could be inferred that the demand for OpenBCI is increasing. Therefore, this study focuses on the feasibility of using OpenBCI in emotion recognition research.}

\begin{figure*}
  \includegraphics[width=\textwidth]{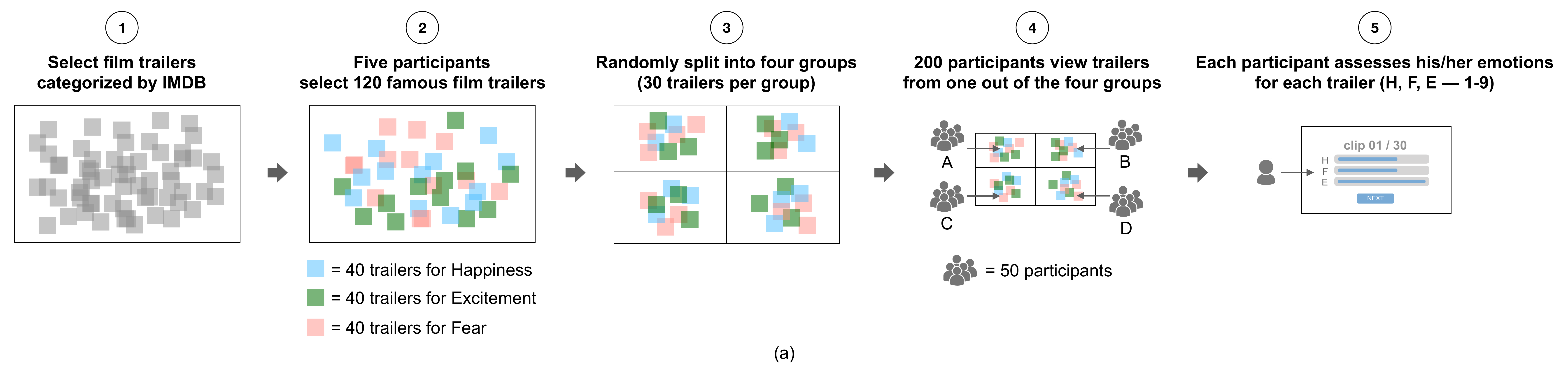}
\end{figure*}
\begin{figure*}
  \vspace*{-0.2in}
  \includegraphics[width=\textwidth]{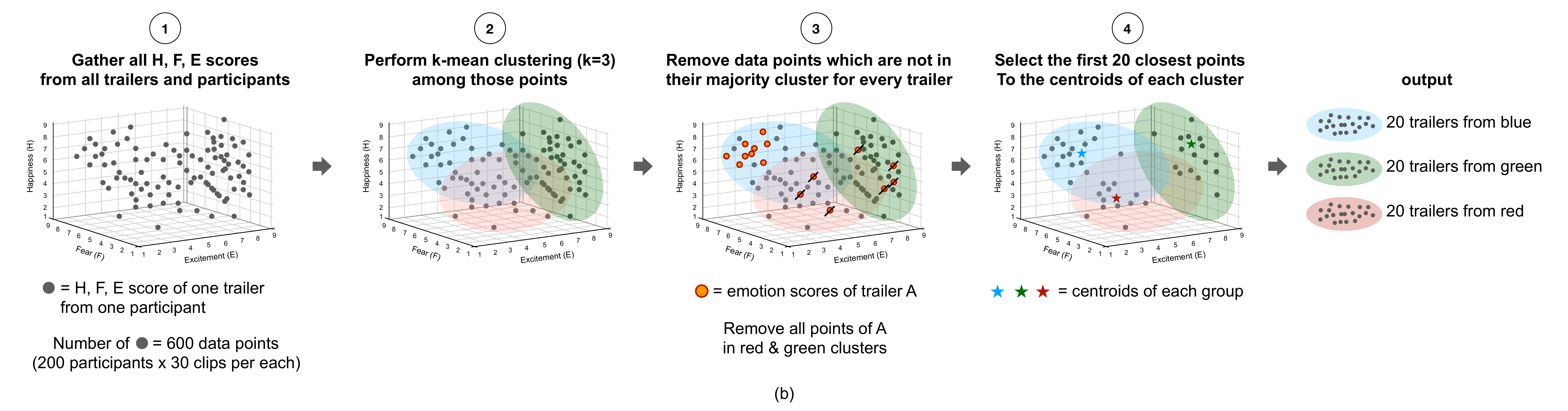}
  \caption{Experiment I starts with (a) random selection of 120 IMDb movie trailers by five individuals. Two-hundred participants are then presented 30 trailers each and afterward asked to assess how the trailer makes them feel. (b) K-means clustering is performed to identify trailers that are effective at inducing emotions in viewers.}
\label{protocol1}
\end{figure*}

\section{Methodology}
This section begins with an experiment to select effective film trailers for emotion elicitation (Experiment I). Consumer grade EEG and peripheral physiological sensors are introduced in Experiment II. The selected trailers from Experiment I are used to invoke emotions, and corresponding EEG and peripheral physiological signals are recorded and stored for subsequent analysis. All experiments fully abide by the 1975 Declaration of Helsinki (revised in 2000), and have been granted ethical approval by the Internal Review Board of Chiang Mai University, Thailand.

\subsection{Experiment I: Affective Video Selection}

Internet  Movie  Database  (IMDb) provides categorical labels for film trailers based on the main genres including drama, comedy, romance, action, sci-fi, horror, and mystery. We simplified these categories into three umbrellas: Happiness, Excitement, and Fear. The selection procedure as illustrated in Figure 1 was followed to identify trailers with explicit emotional content. Five participants randomly picked 40 mainstream film trailers per genre, to give 120 trailer clips in total. The trailer clips contained English soundtrack with native language subtitles (Thai). Afterward, the 120 clips were randomly divided into four groups of 30 clips each. The four groups of 50 participants (n = 200) with near equal numbers of males and females, ranging from 15--22 years old, were then assigned to watch one of the trailer groups. To deal with varying video duration, only the final minute was used for the experiment. At the end of each clip, the participants were required to assess their experienced emotions through a qualitative measure, specifically by choosing a number along a continuous scale of 1--9 that they felt best represented their personal levels of Valence (V), Arousal (A), Happiness (H), Fear (F), and Excitement (E).

To analyze the qualitative data reported by all participants, we attempted two standard clustering methods -- i.e., K-means clustering and Gaussian Mixture Model (GMM) \textbf{}. H, F, and E scores were features that corresponded to distinctive emotional states. We then used Davies-Bouldin index (DB-index) as a metric to evaluate the models providing K=(2, 3, 4, ...,10). Further, we applied the Elbow method to locate the optimal K; K-means clustering with 3 clusters was chosen. Since the same clips were scored by different participants, it was possible that they might not be allocated to the same cluster. Hence, we used only the data points that belonged to their majority class (as inferred from the calculated mode value) and filtered out the rest. All remaining data points from the same clip were averaged, leaving only a single representative point per clip. Finally, we calculated a Euclidean distance of each point from the cluster centroid. In order to obtain the clips that best represented their classes, 20 points closest to the cluster centroid were selected which added up to a total of 60 clips.

\begin{figure}
\centering
\includegraphics[width=0.35\textwidth]{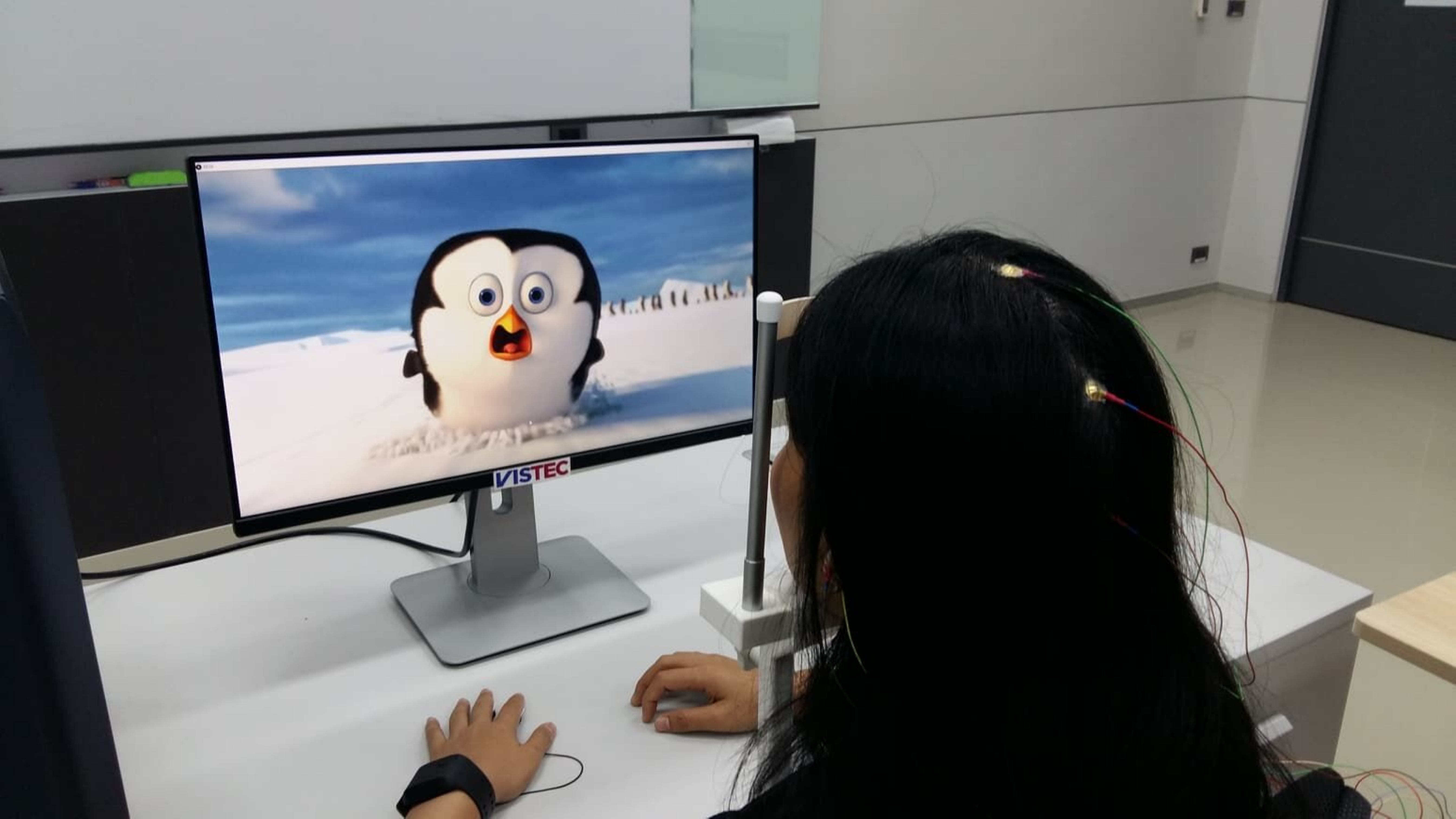}
\caption[]{Experimental setup.}
\label{setup}
\end{figure}

\begin{figure*}
  \includegraphics[width=\textwidth]{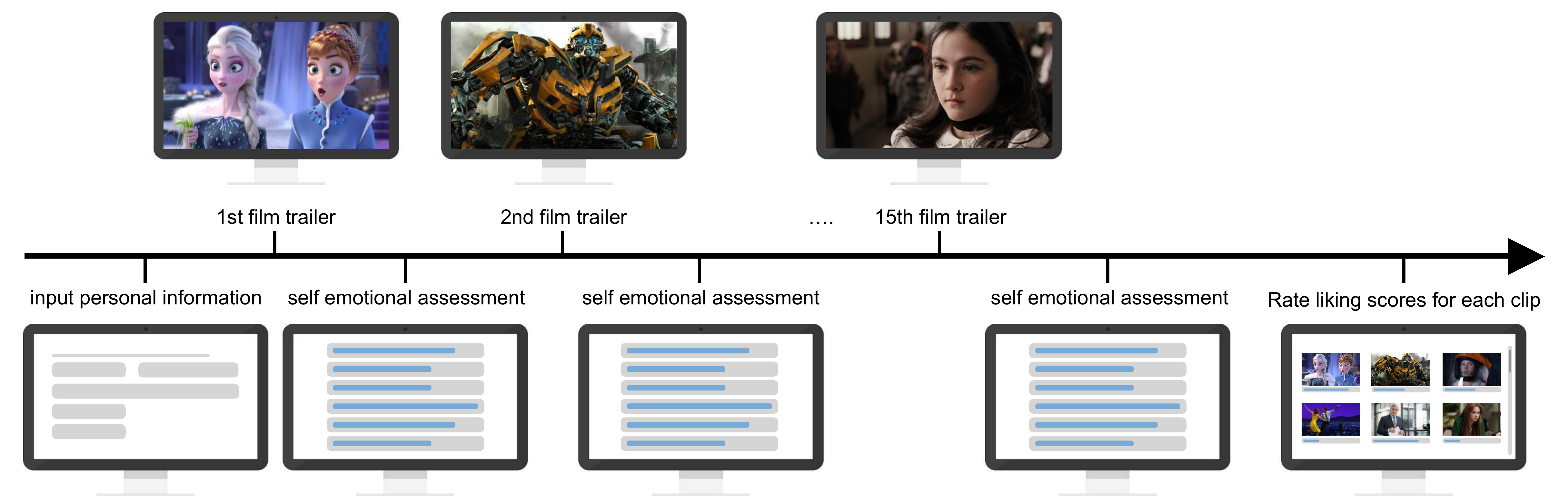}
  \caption{Each experimental run begins with the acquisition of demographic data including participant’s personal information, hours of sleep, level of tiredness, and favorite movie genre. Video clips are played in full-screen mode and a display of emotion self-assessment form is launched at the end of each clip. Screenshots shown here are of six movies by the titles in \cite{frozen2013, transformers2017,orphan2009,lalaland2016,the_intern2015,arrival2016}}
\label{protocol}
\end{figure*}

\subsection{Experiment II: Emotion Recognition Using OpenBCI}
\subsubsection{Sensors}

We used OpenBCI to record emotion-related EEG signals. OpenBCI is the only consumer grade EEG on the market that offers open-source code (software) and hardware (schematic). The advantage of an open-source device is that it directly allows the development of real-time research applications via typical programming languages without additional charges. Moreover, electrode placement is flexible when compared to other devices. To cover brain wave information across the scalp, an eight-channel EEG electrode montage was selected ($Fp_1$, $Fp_2$, $F_z$, $C_z$, $T_3$, $T_4$, $P_z$, and $O_z$) with reference and ground on both mastoids. This montage is a subset of the international standard 10--20 system for EEG electrode placement. In addition to EEG, we also recorded peripheral physiological signals from a wearable Empatica4 or E4 device \cite{Empatica}, which is equipped with multiple sensors, including electrodermal activity (EDA), skin temperature (Temp), and blood volume pulse (BVP) with their derivatives (heart rate, inter-beat interval, etc.). Interestingly, E4 was recently utilized by a research team to record electrodermal activity (EDA) in 25 healthy participants in the study of arousal and valence recognition \cite{sensorJ1}.

\subsubsection{Participants}

Forty-three healthy individuals aged between 16 and 34 (21 male and 22 female) were recruited as volunteers for the study. They were fitted with two monitoring devices for simultaneous recording: OpenBCI (EEG) and Empatica4 (E4). E4 was strapped on the wrist with extension cables and gel electrodes attached (Kendall Foam Electrodes) to measure EDA on the middle knuckle of the index and middle finger. During the main experiment, the participants were instructed to place their chin on the designated chin rest and stay as still as possible in order to minimize artifact interference. We also strove to maintain the impedance below 5 k$\Omega$ for EEG. The experimental setup is depicted in \autoref{setup}.

\subsubsection{Experimental Protocol}

We developed a software to present the emotion-inducing video clips sequentially, and recorded synchronous EEG and peripheral physiological signals with two wearable devices (see \autoref{setup}). Although we intended to elicit a specific emotion from the clip as defined by IMDb tags, each participant might have experienced emotions that were incongruent with the expectation. Owing to this, when labelling the data, each participant was instructed to rate his/her own emotional reaction to the clip while being blinded to its IMDb tags. To avoid potential misunderstanding, before commencing the main experiment, we methodically described the procedures and the meaning of every emotion score. Additionally, an example clip, which was not used in the actual experiment, was shown during a mock trial to test the participants' understanding. We then presented a questionnaire for them to rate their experienced emotions pertaining to Arousal, Valence, Happiness, Fear, and Excitement, on a scale of 1 to 9. Afterward, we checked their answers to ensure that they understood correctly. Subsequently the actual experiment was launched, as shown in \autoref{protocol}. Firstly, we collected personal information (age, gender, hours of sleep, level of tiredness, and favorite movie genre). Secondly, a film trailer was played for 1 minute. Then, we presented a screen displaying the self-assessment questionnaire similar to one featured in the trial run. Finally, another film trailer was played, followed by the questionnaire and the process repeated for all 15 clips. More specifically, a fixed set of 9 (out of 15) clips were played to all 43 participants; these included the top three clips from each cluster from Experiment I. The other six video clips were selected randomly: two per cluster.

\subsubsection{Feature Extraction}
In order to obtain stable and pertinent emotional responses from the EEG and peripheral physiological signals, we started the recording after each movie clip had been played for two seconds and stopped two seconds before the clip ended. This means that on the whole, 56 seconds of signals were collected for a single video clip. Typical EEG pre-processing steps were applied including notch filtering using iirnotch, common average reference (CAR) to find the reference signal for all electrode channels, and independent component analysis (ICA) to remove artifact components. Even if there was no EOG present, ICA components were subject to inspection. On average, zero to one component with characteristics most similar to those of EOG was removed in keeping with the manual provided by MNE \cite{mne-man}. Both CAR and ICA were implemented using MNE-python package \cite{mne}. Conventional feature extraction, as shown in Table II, was computed from pre-processed EEG, EDA, BVP, and Temp. These features were based on previous works \cite{deap,mahnob,dreamer} and included as the baseline for comparison to the results of this study. To extract features within a specified range of EEG frequencies, we applied lfilter (from SciPy package) for bandpass filtering \cite{scipy}.

\begin{table}
\caption{Description of features that are extracted from EEG and Empatica signals.}
\centering 
\begin{tabular}{p{1.8cm} p{6cm}} 
\toprule 
\toprule 
Signal & Extracted Features \\
\midrule 

\textbf{EEG}(32) & $\theta$ (3--7 [Hz]), $\alpha$ (8--13 [Hz]), $\beta$ (14--29 [Hz]) and $\gamma$ (30--47 [Hz]) power spectral density for each channel \\\\

\textbf{EDA}(21) & average skin resistance, average of derivative, average of derivative for negative values only, proportion of negative samples in the derivative vs all samples, number of local minima, average rising time, 14 spectral power in the 0--2.4 [Hz] bands, zero crossing rate of skin conductance slow response 0--0.2 [Hz], zero crossing rate of skin conductance very slow response 0--0.08 [Hz] \\\\

\textbf{BVP}(13) & average and standard deviation of HR, HRV, and inter beat intervals, energy ratio between the frequency bands 0.04--0.15 [Hz] and 0.15--0.5 [Hz], spectral power in the bands 0.1--0.2 [Hz], 0.2--0.3 [Hz], 0.3--0.4 [Hz], low frequency 0.01--0.08 [Hz], medium frequency 0.08--0.15 [Hz] and high frequency 0.15--0.5 [Hz] components of HRV power spectrum \\\\

\textbf{Temp}(4) & average, average of its derivative, spectral power in the bands 0--0.1 [Hz] and 0.1--0.2 [Hz] \\

\bottomrule
\bottomrule
\end{tabular}
 
\label{table feature}
\end{table}

\begin{figure*}
\centering
\includegraphics[width=0.7\textwidth]{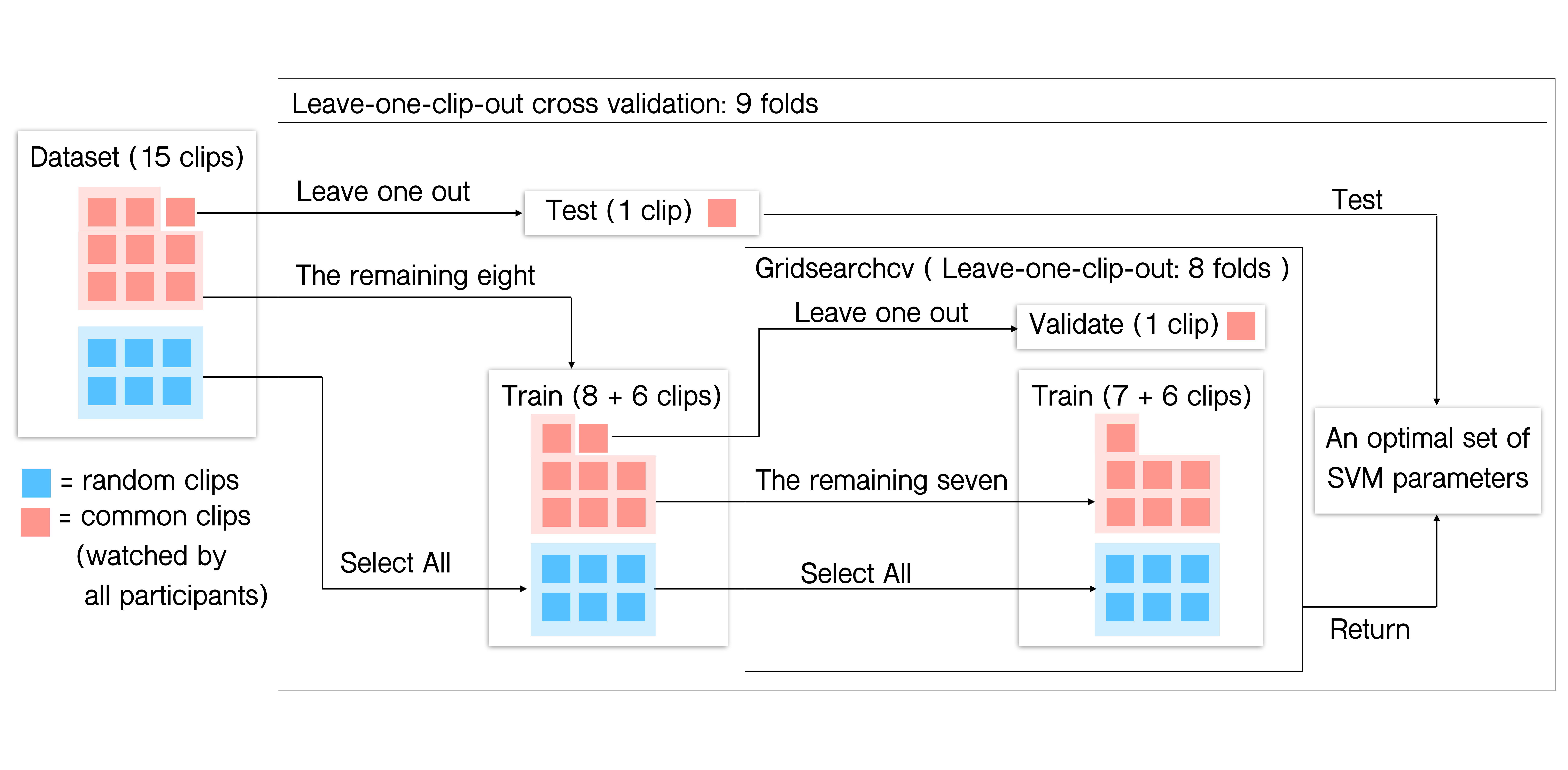}
\caption[]{Schematic of leave-one-clip-out cross-validation steps with gridsearchcv.}
\label{fig_cv}
\end{figure*}

\subsubsection{Emotion Recognition}

For this part, the investigation was divided into three subtasks, as detailed in the bullet points below. All analyses were based on binary classification (low-high valence and arousal). The first analysis relied on a straightforward threshold setting for labeling of two separable classes, while the second and third relied on the use of K-means clustering technique. One sample point correlated to a recording from one participant in response to the short video clip; there were 645 (43 participants $\times$ 15 clips) sample points overall.

\begin{itemize}
\item \textbf{Low-High: Valence/Arousal with Threshold}

We labeled each sample on the basis of associated V and A scores and manually set the threshold at 4.5 (midpoint between 1 to 9) for binary classification. In essence, a sample with V or A score lower than 4.5 was labeled as low V or A, and vice versa. Two binary classification tasks were then carried out, high vs low valence and high vs low arousal. \autoref{table feature} displays the set of input features for the model, the total being 70 features.

\item \textbf{Low-High: Valence/Arousal with Clustering}

We conducted GMM and K-means clustering using V and A scores from the participants in order to select the most suitable clustering method. DB-index was used as a metric and it turned out that K-means clustering with 4 clusters was the best one. According to four combinations of V and A models in a conventional emotion-related study (low (L) or high (H) V and A scores), we labeled samples in the four groups stratified by K-means with LVLA, LVHA, HVLA, and LVHA. The input features were identical to the previous subsection.

\item \textbf{EEG Electrode Channels and Frequency Bands}

In this subtask, the electrode setups were examined for prospective design of a user-friendly EEG device capable of emotion recognition. Here, 3 or 4 out of 8 electrode channels were strategically chosen to explore options for hardware design. The process began by studying the sagittal line of the head ($F_z$, $C_z$, $P_z$ $O_z$), and another eight sets of channels were then created ($FP_1$, $FP_2$, $F_z$), ($FP_1$, $FP_2$, $C_z$), ($FP_1$, $FP_2$, $C_z$), ($FP_1$, $FP_2$, $P_z$), ($FP_1$, $FP_2$, $O_z$), ($T_3$, $T_4$, $F_z$), ($T_3$, $T_4$, $C_z$), ($T_3$, $T_4$, $P_z$) and ($T_3$, $T_4$, $O_z$). These were taken from a combination of either the temporal or frontal lobe and one channel from the sagittal line. In addition, we sought to pinpoint the important frequency bands within the EEG signals, provided the four fundamental bands: $\theta$ (3--7 [Hz]), $\alpha$ (8--13 [Hz]), $\beta$ (14--29 [Hz]), and $\gamma$ (30--47 [Hz]). For binary classification (low-high V and A), we performed labeling in a similar manner to \textit{Low-High: Valence/Arousal with Clustering} and compared the results among the sets of channels and frequency bands.

\end{itemize}

We implemented leave-one-clip-out cross-validation using Support Vector Machine (SVM), as illustrated in \autoref{fig_cv}. Since a fixed set of 9 video clips were commonly seen by the participants (of all 15 videos watched by each), nine-fold cross-validation was conducted. In each fold, one clip was set aside as a test sample and the rest were used as a training set. We normalized all features by using MinMaxScaler to scale them into a common range \cite{scikit-learn}. In the training session, the training set was used to determine the optimal parameters that drive the SVM model to return the best $F_{1}$ score; this was done with \textit{gridsearchcv}. Within \textit{gridsearchcv}, there was also leave-one-clip-out cross-validation. The parameters included kernel (linear, poly, rbf, sigmoid), C (1, 10, 100), degree (3, 4, 5), coef0 (0, 0.01, 0.1), and regularization (L1, L2). Finally, we evaluated the model of each fold by setting the optimal SVM parameters and performed prediction on the test set.

\section{Results}

The result descriptions are sequentially organized according to the order of the experiments. In Experiment I, the output from K-means clustering facilitated the sorting of emotion-eliciting video clips into appropriate groups for use in Experiment II. The resulting datasets were analyzed with a simple machine learning algorithm and the outputs were contrasted to relevant datasets from previous studies on affective computing. The performance appeared to be on par if not better when compared to the existing works employing similar emotion-eliciting method and emotion recognition algorithm.

\begin{table}
\centering
\scriptsize
\caption{Clustering methods and associated Davies-Bouldin indices}
\centering 
\begin{threeparttable}
\begin{tabular}{p{2.0cm} c} 
\toprule 
\toprule 
Method & Davies-Bouldin Score  \\
\midrule 
\textbf{GMM:}  & \\
\hspace{4mm} 2 cluster & 1.4816 \\
\hspace{4mm} 3 cluster & 3.9107 \\
\hspace{4mm} 4 cluster & 2.2370 \\
\hspace{4mm} 5 cluster & 1.3389 \\
\hspace{4mm} 6 cluster & 4.7667 \\
\textbf{K-Means:} &  \\
\hspace{4mm} 2 cluster & 1.1771 \\
\hspace{4mm} \textbf{3 cluster}  & \textbf{0.8729} \\
\hspace{4mm} 4 cluster & 0.8866 \\
\hspace{4mm} 5 cluster & 0.8956 \\
\hspace{4mm} 6 cluster & 0.8842 \\
\bottomrule
\bottomrule
\end{tabular}
 \begin{tablenotes}
\small \item \textit{Optimum shown in boldface}
\end{tablenotes}
\end{threeparttable}
\label{table freq}
\label{tableDBindex1}
\end{table}

\begin{figure}
\centering
\includegraphics[width=0.45\textwidth]{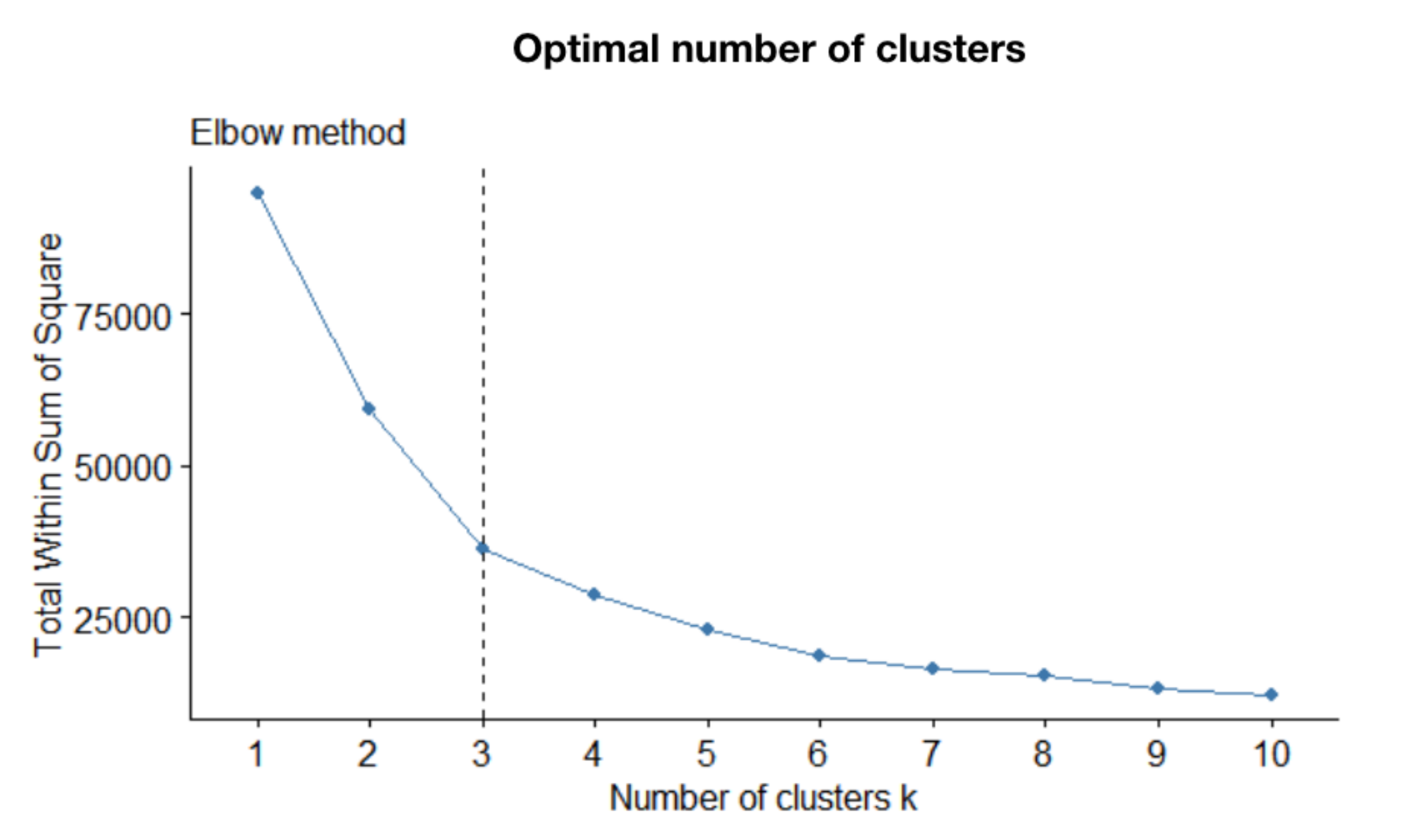}
\caption[]{Elbow method to determine the optimal number of K-means clusters for affective video grouping, K=3}
\label{figElbow1}
\end{figure}

\begin{figure}
\centering
\begin{subfigure}{3.5cm}
  \centering
  \includegraphics[width=1\textwidth]{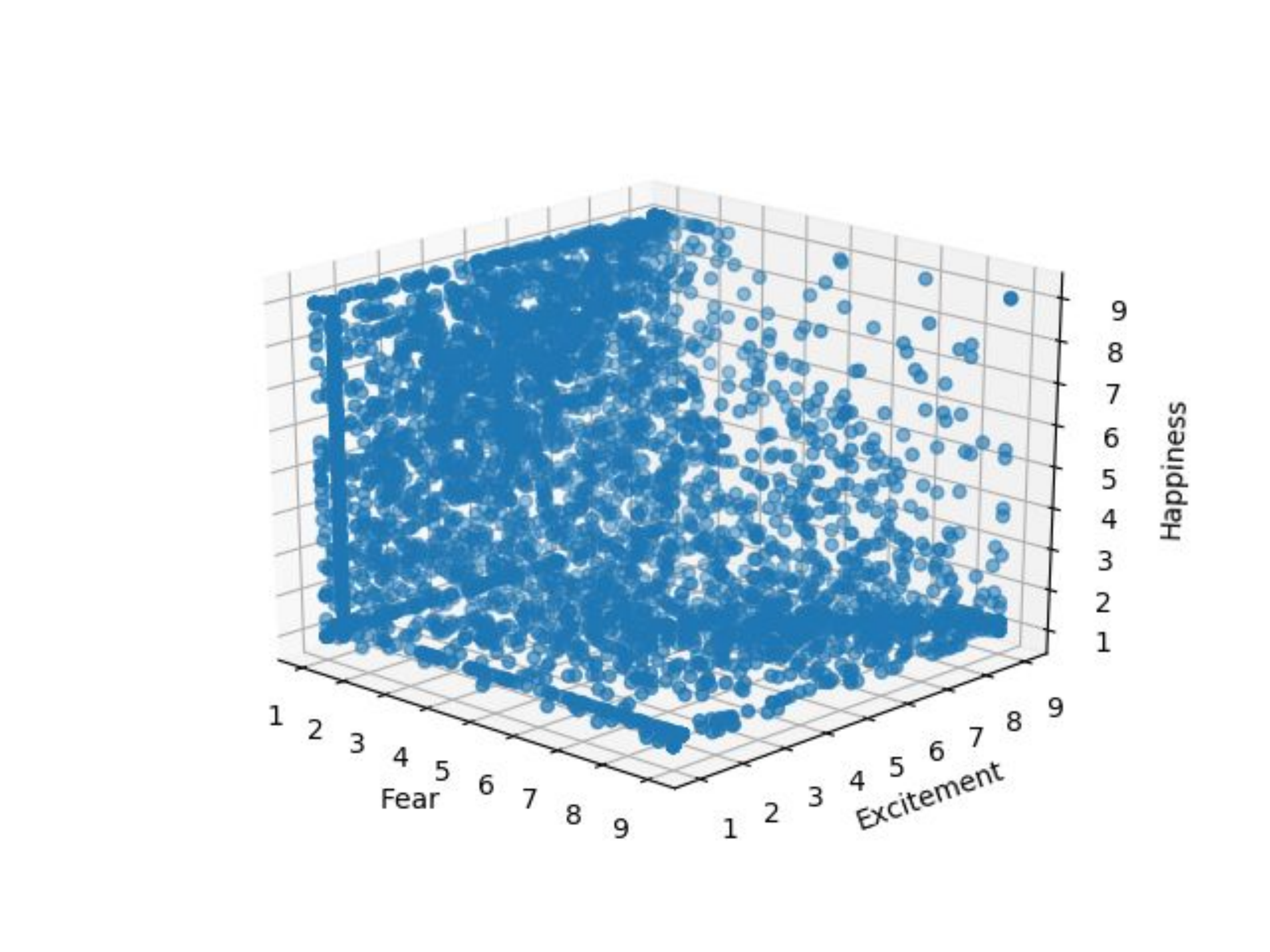}
  \caption{}
  \label{fig4:a}
\end{subfigure}
\begin{subfigure}{3.5cm}
  \centering
  \includegraphics[width=1\textwidth]{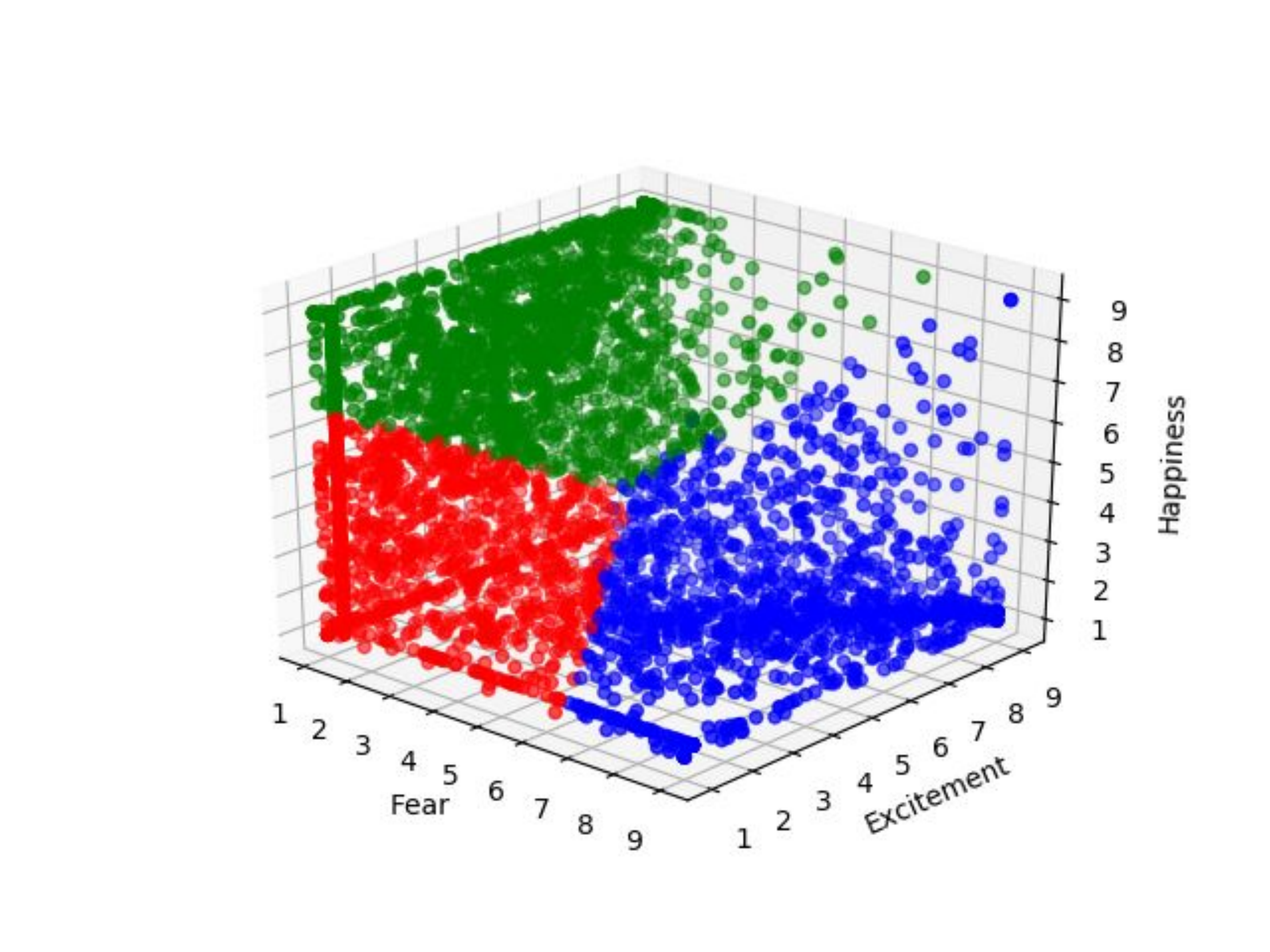}
 \caption{}
  \label{fig4:b}
\end{subfigure}
\begin{subfigure}{3.5cm}
  \centering
  \includegraphics[width=1\textwidth]{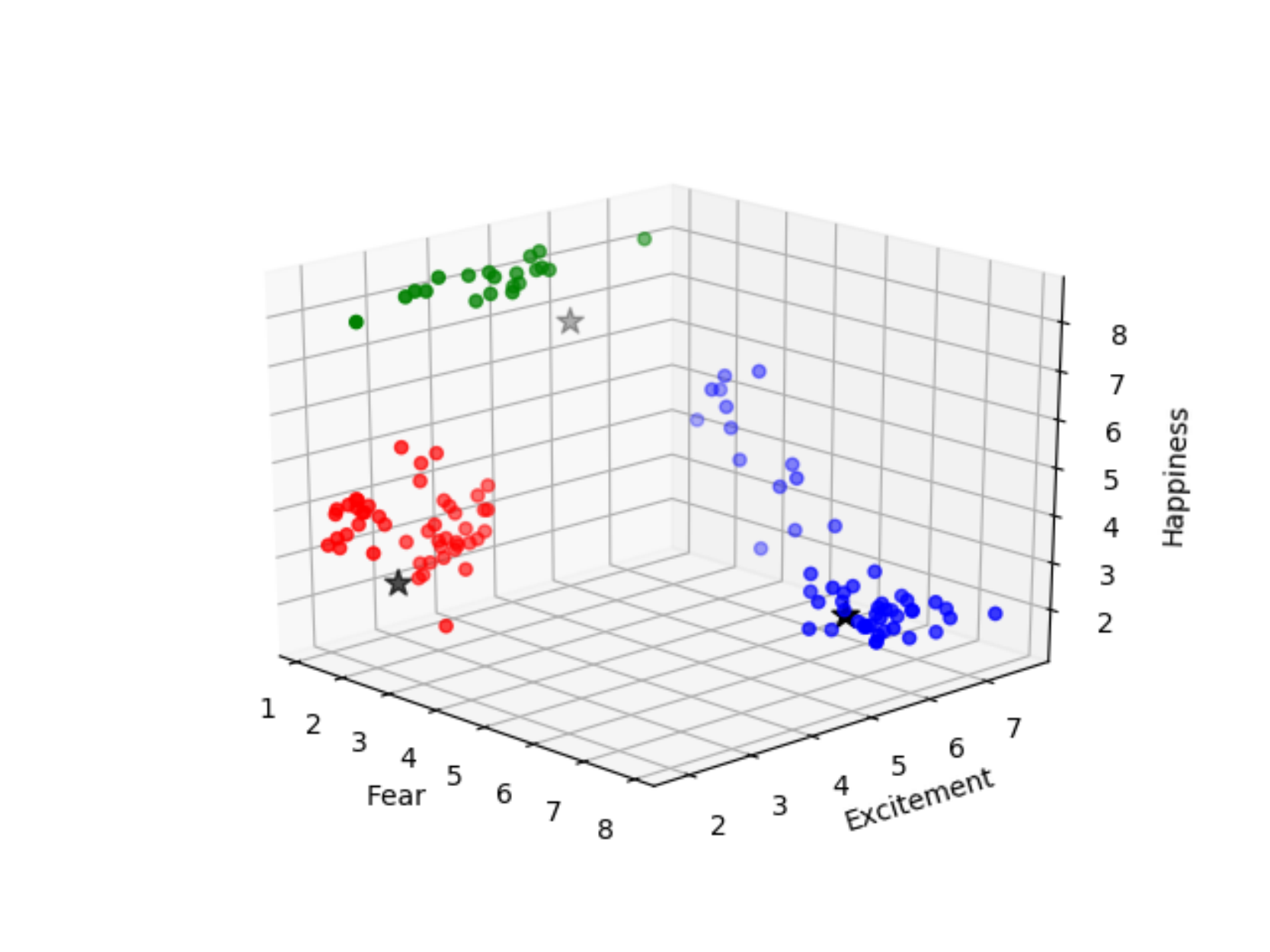}
  \caption{}
  \label{fig4:c}
\end{subfigure}
\begin{subfigure}{3.5cm}
  \centering
  \includegraphics[width=1\textwidth]{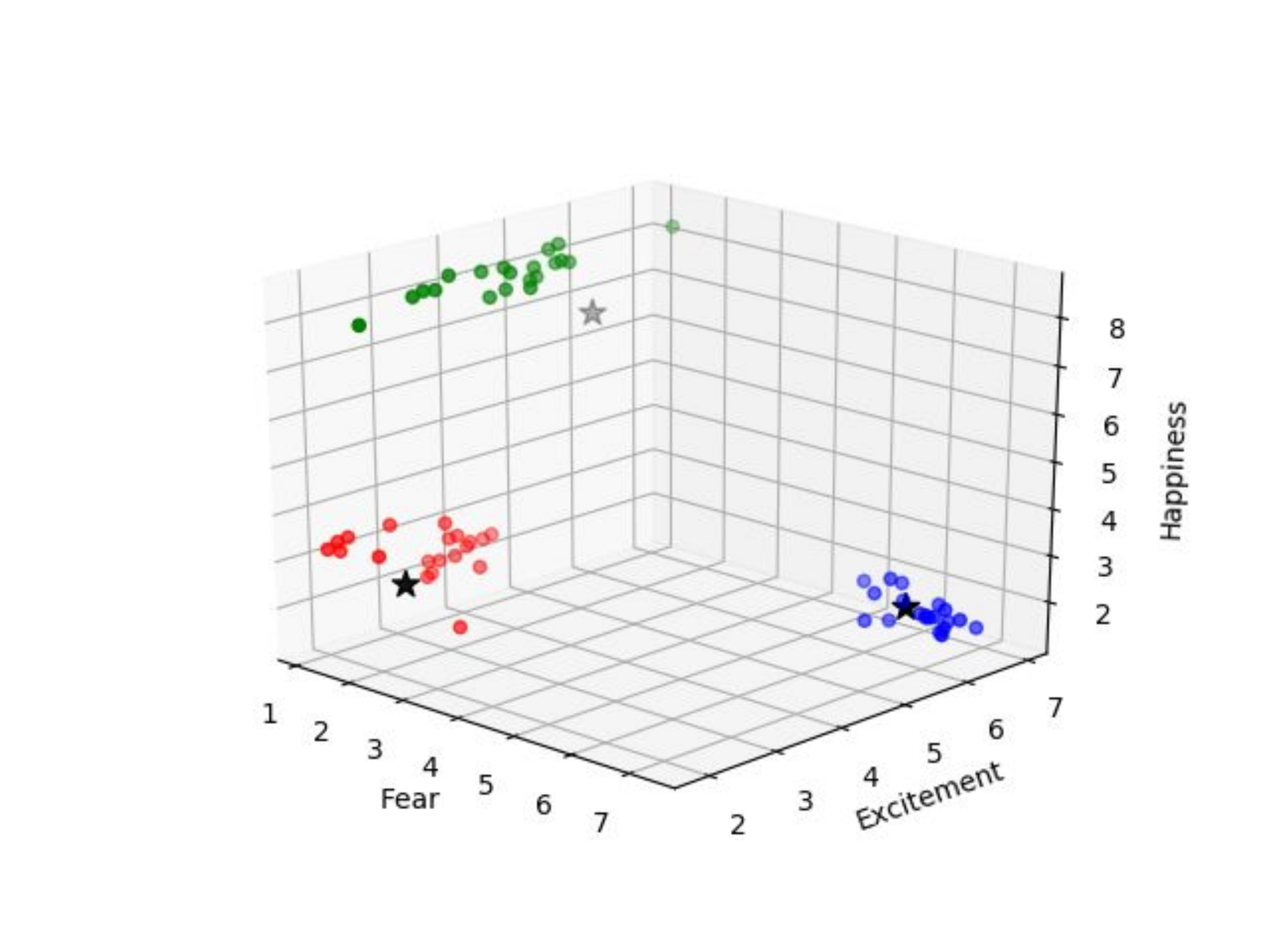}
  \caption{}
  \label{fig4:d}
\end{subfigure}
\caption{(a) Scatter plot of all qualitative samples from Experiment I. K-means clustering on (a) yields the output clusters as shown in (b). After removing the data points that do not belong to the majority class (as inferred from the mode value), the remainder is shown in (c). In (d), 20  points per cluster are retained to figure out the nearest distances to the centroids for emotion elicitation in Experiment II.}

\label{fig4} 
\end{figure}

\begin{table*}
\caption{Comprehensive list of 60 affective video titles featured in Experiment II to allow monitoring of emotionally relevant EEG and peripheral physiological signals. Means and standard deviations of quantified attributes (Valence, Arousal, Happiness, Fear, Excitement) are listed in adjacent columns. 'Samp' (short for sample) denotes the number of participants having seen the particular clip and submitted subjective ratings.}
\label{tableclip}
\scriptsize
\centering
\begin{tabular}{lllllllll} 
\toprule
\textbf{ID} & \textbf{Movie Title}    & \textbf{Affective Tags (IMDb)}             & \textbf{Valence} & \textbf{Arousal} & \textbf{Happiness} & \textbf{Fear} & \textbf{Excitement} & \textbf{Samp}  \\ 
\hline
1           & The Help                       & Drama                                      & 4.28 $\pm$ 2.08      & 3.34 $\pm$ 1.96      & 4.41 $\pm$ 2.12        & 1.26 $\pm$ 0.37   & 2.56 $\pm$ 1.66         & 82                        \\
2           & Star Wars: The Last Jedi       & Action,~Adventure,~Fantasy,~Sci-Fi         & 4.98 $\pm$ 2.15      & 4.36 $\pm$ 2.08      & 4.06 $\pm$ 2.16        & 1.79 $\pm$ 1.00   & 5.09 $\pm$ 2.12         & 81                        \\
3           & Suicide Squad                  & Action,~Adventure,~Fantasy,~Sci-Fi         & 4.73 $\pm$ 2.02      & 4.12 $\pm$1.90       & 4.36 $\pm$ 2.18        & 1.73 $\pm$ 1.41   & 4.37 $\pm$ 2.14         & 81                        \\
4           & Pacific Rim                    & Action,~Adventure,~Sci-Fi                  & 4.10 $\pm$ 2.03      & 3.81 $\pm$ 2.16      & 3.93 $\pm$ 2.04        & 1.31 $\pm$ 0.52   & 4.25 $\pm$ 2.12         & 44                        \\
5           & War for the Planet of the Apes & Action, Adventure, Drama, Sci-Fi, Thriller & 4.73 $\pm$ 2.06      & 4.04 $\pm$ 2.20      & 2.57 $\pm$ 1.90        & 2.67 $\pm$ 1.76   & 4.83 $\pm$ 2.08         & 41                        \\
6           & God Help the Girl              & Drama, Musical,~Romance                    & 4.38 $\pm$ 2.45      & 3.09 $\pm$ 2.02      & 5.00 $\pm$ 2.45        & 1.27 $\pm$ 0.47   & 3.01 $\pm$ 1.85         & 43                        \\
7           & Rogue One: A Star Wars Story   & Action, Adventure, Sci-Fi                  & 5.09 $\pm$ 1.78      & 4.65 $\pm$ 2.15      & 3.34 $\pm$ 2.17        & 2.04 $\pm$ 1.23   & 5.28 $\pm$ 2.27         & 42                        \\
8           & Blade Runner 2049              & Drama, Mystery, Sci-Fi                     & 4.44 $\pm$ 2.47      & 4.34 $\pm$ 2.38      & 3.15 $\pm$ 2.02        & 2.39 $\pm$ 1.73   & 4.81 $\pm$ 2.29         & 44                        \\
9           & Hope Springs                   & Comedy, Drama, Romance                     & 4.78 $\pm$ 2.46      & 3.56 $\pm$ 2.12      & 5.09 $\pm$ 2.34        & 1.19 $\pm$ 0.17   & 2.84 $\pm$ 1.89         & 45                        \\
10          & Ghost in the Shell             & Action, Drama, Sci-Fi                      & 4.77 $\pm$ 2.10      & 4.28 $\pm$ 2.27      & 3.41 $\pm$ 2.13        & 2.07 $\pm$ 1.57   & 5.01 $\pm$ 2.27         & 47                        \\
11          & Point Break                    & Action, Crime, Sport                       & 4.65 $\pm$ 2.33      & 4.49 $\pm$ 2.40      & 3.31 $\pm$ 2.33        & 2.08 $\pm$ 1.41   & 5.12 $\pm$ 2.56         & 40                        \\
12          & The Hunger Games               & Adventure, Sci-Fi, Thriller                & 5.42 $\pm$ 2.26      & 4.76 $\pm$ 2.18      & 3.66 $\pm$ 2.09        & 2.06 $\pm$ 1.34   & 5.38 $\pm$ 2.05         & 42                        \\
13          & Crazy, Stupid, Love.           & Comedy, Drama, Romance                     & 4.82 $\pm$ 2.54      & 4.01 $\pm$ 2.52      & 5.12 $\pm$ 2.64        & 1.21 $\pm$ 0.25   & 3.06 $\pm$ 2.35         & 43                        \\
14          & Arrival                        & Drama, Mystery, Sci-F                      & 4.84 $\pm$ 2.22      & 4.82 $\pm$ 2.28      & 3.28 $\pm$ 2.03        & 2.64 $\pm$ 1.90   & 5.66 $\pm$ 2.40         & 44                        \\
15          & Mr. Hurt                       & Comedy, Romance                            & 4.50 $\pm$ 2.21      & 3.59 $\pm$ 2.20      & 4.66 $\pm$ 2.19        & 1.23 $\pm$ 0.21   & 2.41 $\pm$ 1.43         & 42                        \\
16          & American Assassin              & Action, Thriller                           & 4.19 $\pm$ 2.18      & 4.80 $\pm$ 2.33      & 2.90 $\pm$ 1.88        & 2.48 $\pm$ 1.92   & 5.03 $\pm$ 2.36         & 43                        \\
17          & G.I. Joe: Retaliation          & Action, Adventure, Sci-Fi                  & 4.69 $\pm$ 2.39      & 4.11 $\pm$ 2.31      & 3.46 $\pm$ 2.09        & 1.55 $\pm$ 0.93   & 5.02 $\pm$ 2.54         & 48                        \\
18          & Beginners                      & Comedy, Drama, Romance                     & 4.42 $\pm$ 2.45      & 3.08 $\pm$ 2.18      & 4.97 $\pm$ 2.20        & 1.23 $\pm$ 0.27   & 2.42 $\pm$ 1.48         & 44                        \\
19          & Open Grave                     & Horror, Mystery, Thriller                  & 3.70 $\pm$ 1.90      & 4.70 $\pm$ 2.30      & 1.90 $\pm$ 1.14        & 4.03 $\pm$ 2.40   & 4.55 $\pm$ 2.46         & 43                        \\
20          & Flipped                        & Comedy, Drama, Romance                     & 5.05 $\pm$ 2.69      & 3.75 $\pm$ 2.40      & 5.43 $\pm$ 2.48        & 1.16 $\pm$ 0.08   & 2.88 $\pm$ 2.12         & 44                        \\
21          & The Choice                     & Drama, Romance                             & 4.58 $\pm$ 2.11      & 4.12 $\pm$ 1.98      & 4.63 $\pm$ 2.16        & 1.43 $\pm$ 0.68   & 3.24. $\pm$ 1.71        & 82                        \\
22          & Danny Collins                  & Biography, Comedy, Drama                   & 4.54 $\pm$ 2.20      & 3.55 $\pm$ 2.07      & 5.01 $\pm$ 2.09        & 1.17 $\pm$ 0.09   & 2.76 $\pm$ 1.68         & 82                        \\
23          & The Big Sick                   & Comedy, Drama, Romance                     & 4.44 $\pm$ 2.10      & 3.21 $\pm$ 1.80      & 4.56 $\pm$ 2.15        & 1.28 $\pm$ 0.35   & 2.93 $\pm$ 1.90         & 86                        \\
24          & Monsters University            & Animation, Adventure, Comedy               & 5.55 $\pm$ 2.07      & 4.01 $\pm$ 2.12      & 6.15 $\pm$ 2.12        & 1.24 $\pm$ 0.23   & 4.44 $\pm$ 2.14         & 47                        \\
25          & Kung Fu Panda 3                & Animation, Action, Adventure               & 6.11 $\pm$ 2.17      & 4.46 $\pm$ 2.31      & 6.44 $\pm$ 2.37        & 1.25 $\pm$ 0.24   & 4.14 $\pm$ 2.11         & 47                        \\
26          & Baby Driver                    & Action, Crime, Drama                       & 5.29 $\pm$ 2.10      & 4.65 $\pm$ 2.27      & 5.19 $\pm$ 2.32        & 1.53 $\pm$ 0.88   & 5.30 $\pm$ 2.28         & 44                        \\
27          & The Good Dinosaur              & Animation, Adventure, Comedy               & 6.43 $\pm$ 2.15      & 4.63 $\pm$ 2.33      & 6.47 $\pm$ 2.18        & 1.19 $\pm$ 0.41   & 4.22 $\pm$ 2.07         & 43                        \\
28          & About Time                     & Comedy, Drama, Fantasy                     & 5.25 $\pm$ 2.60      & 4.29 $\pm$ 2.68      & 5.97 $\pm$ 2.32        & 1.22 $\pm$ 0.34   & 4.25 $\pm$ 2.24         & 43                        \\
29          & Ordinary World                 & Comedy, Drama, Music                       & 4.88 $\pm$ 1.72      & 3.93 $\pm$ 1.72      & 5.47 $\pm$ 1.68        & 1.22 $\pm$ 0.20   & 3.45 $\pm$ 1.98         & 49                        \\
30          & Lion                           & Biography, Drama                           & 5.36 $\pm$ 2.30      & 4.62 $\pm$ 2.65      & 5.01 $\pm$ 2.49        & 1.79 $\pm$ 1.25   & 3.93 $\pm$ 2.35         & 48                        \\
31          & Shrek Forever After            & Animation, Adventure, Comedy               & 5.87 $\pm$ 2.12      & 3.85 $\pm$ 2.37      & 6.29 $\pm$ 2.29        & 1.26 $\pm$ 0.29   & 4.35 $\pm$ 2.49         & 44                        \\
32          & Chappie                        & Action, Crime, Drama                       & 5.47 $\pm$ 2.26      & 4.43 $\pm$ 2.20      & 4.54 $\pm$ 2.45        & 1.69 $\pm$ 0.73   & 5.31 $\pm$ 2.31         & 47                        \\
33          & Guardians of the Galaxy Vol. 2 & Action, Adventure, Sci-Fi                  & 6.15 $\pm$ 2.40      & 4.61 $\pm$ 2.34      & 5.85 $\pm$ 2.40        & 1.44 $\pm$ 1.01   & 5.56 $\pm$ 2.50         & 48                        \\
34          & The Intern                     & Comedy, Drama                              & 6.34 $\pm$ 2.02      & 5.06 $\pm$ 2.13      & 6.31 $\pm$ 1.98        & 1.23 $\pm$ 0.39   & 3.74 $\pm$ 2.34         & 42                        \\
35          & La La Land                     & Comedy, Drama, Music                       & 5.44 $\pm$ 2.24      & 4.09 $\pm$ 2.37      & 5.55 $\pm$ 2.29        & 1.49 $\pm$ 1.28   & 3.15 $\pm$ 1.99         & 47                        \\
36          & Ice Age: Collision Course      & Animation, Adventure, Comedy               & 6.38 $\pm$ 2.36      & 4.96 $\pm$ 2.40      & 6.92 $\pm$ 1.94        & 1.21 $\pm$ 0.26   & 4.87 $\pm$ 2.36         & 43                        \\
37          & Frozen                         & Animation, Adventure, Comedy               & 5.88 $\pm$ 2.40      & 4.17 $\pm$ 2.38      & 6.31 $\pm$ 2.41        & 1.24 $\pm$ 0.40   & 4.35 $\pm$ 2.36         & 47                        \\
38          & Transformers: The Last Knight  & Action,~Adventure,~Sci-Fi                  & 4.57 $\pm$ 2.06      & 4.18 $\pm$ 1.96      & 4.10 $\pm$ 2.33        & 1.91 $\pm$ 1.22   & 4.87 $\pm$ 1.97         & 42                        \\
39          & Divergent                      & Adventure,~Mystery,~Sci-Fi                 & 5.87 $\pm$ 1.81      & 4.87 $\pm$ 2.00      & 4.75 $\pm$ 2.27        & 1.95 $\pm$ 1.34   & 5.84 $\pm$ 2.05         & 49                        \\
40          & Why Him?                       & Comedy                                     & 5.85 $\pm$ 2.24      & 4.60 $\pm$ 2.40      & 6.03 $\pm$ 2.30        & 1.25 $\pm$ 0.39   & 4.06 $\pm$ 2.44         & 43                        \\
41          & The Boy                        & Horror, Mystery, Thriller                  & 3.85 $\pm$ 2.09      & 4.92 $\pm$ 1.97      & 1.78 $\pm$ 1.06        & 5.21 $\pm$ 2.23   & 5.01 $\pm$ 2.16         & 82                        \\
42          & Jigsaw                         & Crime, Horror, Mystery                     & 4.04 $\pm$ 2.14      & 4.68 $\pm$ 2.15      & 2.07 $\pm$ 1.37        & 4.65 $\pm$ 2.18   & 4.91 $\pm$ 2.18         & 86                        \\
43          & Shutter                        & Horror, Mystery, Thriller                  & 3.53 $\pm$ 2.20      & 4.71 $\pm$ 2.25      & 1.68 $\pm$ 0.97        & 5.23 $\pm$ 2.34   & 4.63 $\pm$ 2.29         & 81                        \\
44          & Ladda Land                     & Horror                                     & 4.61 $\pm$ 2.20      & 4.81 $\pm$ 2.23      & 1.95 $\pm$ 1.47        & 5.62 $\pm$ 2.13   & 4.88 $\pm$ 2.07         & 42                        \\
45          & No One Lives                   & Horror, Thriller                           & 4.20 $\pm$ 2.04      & 4.84 $\pm$ 2.28      & 1.94 $\pm$ 1.35        & 4.97 $\pm$ 2.56   & 5.07 $\pm$ 2.33         & 47                        \\
46          & Tales from the Crypt           & Horror                                     & 3.83 $\pm$ 2.22      & 4.41 $\pm$ 2.21      & 2.21 $\pm$ 1.88        & 4.67 $\pm$ 2.36   & 4.68 $\pm$ 2.41         & 44                        \\
47          & Orphan                         & Horror, Mystery, Thriller                  & 4.07 $\pm$ 2.35      & 5.18 $\pm$ 2.11      & 1.85 $\pm$ 1.40        & 5.11 $\pm$ 2.15   & 4.68 $\pm$ 2.27         & 40                        \\
48          & Unfriended                     & Drama, Horror, Mystery                     & 4.34 $\pm$ 2.57      & 5.40 $\pm$ 2.57      & 1.98 $\pm$ 1.93        & 5.34 $\pm$ 2.53   & 5.37 $\pm$ 2.55         & 43                        \\
49          & Poltergeist                    & Horror, Thriller                           & 4.13 $\pm$ 2.44      & 5.28 $\pm$ 2.55      & 1.91 $\pm$ 1.70        & 5.85 $\pm$ 2.36   & 5.20 $\pm$ 2.60         & 43                        \\
50          & Jeruzalem                      & Horror~                                    & 4.20 $\pm$ 2.12      & 4.82 $\pm$ 2.14      & 2.02 $\pm$ 1.42        & 5.00 $\pm$ 2.23   & 4.90 $\pm$ 2.27         & 49                        \\
51          & Leatherface                    & Crime, Horror, Thriller                    & 3.92 $\pm$ 2.11      & 4.77 $\pm$ 2.39      & 1.89 $\pm$ 1.27        & 4.93 $\pm$ 2.53   & 4.90 $\pm$ 2.50         & 47                        \\
52          & The Babadook                   & Drama, Horror                              & 3.62 $\pm$ 2.06      & 4.84 $\pm$ 2.19      & 1.67 $\pm$ 1.05        & 5.34 $\pm$ 2.11   & 4.74 $\pm$ 2.17         & 43                        \\
53          & Oculus                         & Horror, Mystery                            & 4.12 $\pm$ 2.11      & 5.81 $\pm$ 1.90      & 1.76 $\pm$ 1.29        & 6.07 $\pm$ 1.81   & 5.36 $\pm$ 2.28         & 42                        \\
54          & The Witch                      & Horror, Mystery                            & 3.69 $\pm$ 2.17      & 4.69 $\pm$ 2.44      & 1.89 $\pm$ 1.38        & 5.12 $\pm$ 2.40   & 4.54 $\pm$ 2.37         & 42                        \\
55          & Trick 'r Treat                 & Comedy, Horror, Thriller                   & 4.50 $\pm$ 2.27      & 5.59 $\pm$ 2.30      & 1.93 $\pm$ 1.42        & 5.68 $\pm$ 2.26   & 5.55 $\pm$ 2.34         & 42                        \\
56          & The Woman in Black             & Drama, Fantasy, Horror                     & 4.23 $\pm$ 2.29      & 4.67 $\pm$ 2.27      & 1.94 $\pm$ 1.54        & 5.21 $\pm$ 2.33   & 4.79 $\pm$ 2.16         & 42                        \\
57          & The Possession                 & Horror, Thriller                           & 4.83 $\pm$ 2.47      & 5.32 $\pm$ 2.29      & 1.82 $\pm$ 1.44        & 5.90 $\pm$ 2.42   & 5.66 $\pm$ 2.38         & 42                        \\
58          & Crimson Peak                   & Drama, Fantasy, Horror                     & 3.93 $\pm$ 2.15      & 4.79 $\pm$ 2.33      & 2.33 $\pm$ 1.71        & 4.81 $\pm$ 2.43   & 5.10 $\pm$ 2.16         & 43                        \\
59          & Program na winyan akat         & Horror, Thriller                           & 3.95 $\pm$ 2.11      & 4.98 $\pm$ 2.48      & 1.72 $\pm$ 1.20        & 5.52 $\pm$ 2.45   & 5.00 $\pm$ 2.26         & 43                        \\
60          & The Pact                       & Horror, Mystery, Thriller                  & 4.37 $\pm$ 2.00      & 5.56 $\pm$ 2.40      & 1.86 $\pm$ 1.42        & 6.23 $\pm$ 2.05   & 5.62 $\pm$ 2.50         & 44                        \\
\bottomrule
\end{tabular}
\end{table*}

\begin{table}
\centering
\scriptsize
\caption{Davies-Bouldin indices achieved with GMM and K-means clustering of Valence and Arousal scores.}
\centering 

\begin{threeparttable}
\begin{tabular}{p{2.0cm} c} 
\toprule 
\toprule 
Method & Davies-Bouldin Score  \\
\midrule

\textbf{GMM:}  & \\
\hspace{4mm} 2 clusters & 7.3844 \\
\hspace{4mm} 3 clusters & 7.2030 \\
\hspace{4mm} 4 clusters & 1.6682\\
\hspace{4mm} 5 clusters & 1.2772 \\
\hspace{4mm} 6 clusters & 2.4652\\
\textbf{K-Means:} &  \\
\hspace{4mm} 2 clusters & 0.8216 \\
\hspace{4mm} 3 clusters & 1.0010 \\
\hspace{4mm} \textbf{4 clusters} & \textbf{0.8095} \\
\hspace{4mm} 5 clusters & 0.8556 \\
\hspace{4mm} 6 clusters & 0.8255 \\
\bottomrule
\bottomrule
\end{tabular}
 \begin{tablenotes}
\small \item \textit{Optimum shown in boldface}
\end{tablenotes}
\end{threeparttable}
\label{tableDBindex2}
\end{table}

\begin{figure}
\centering
\begin{subfigure}{4cm}
  \centering
  \includegraphics[width=1\textwidth]{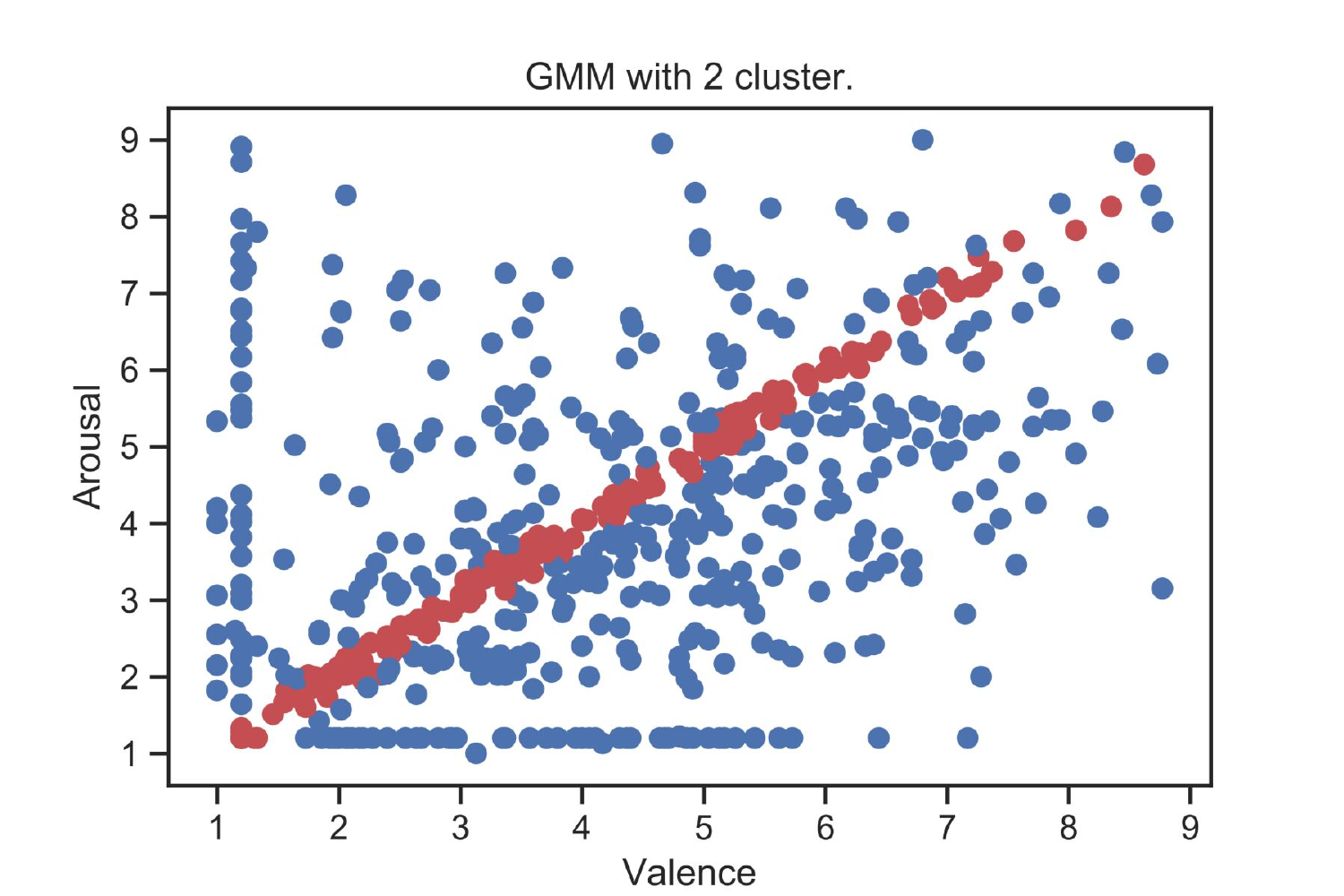}
  \caption{}
  \label{fig:c}
\end{subfigure}
\begin{subfigure}{4cm}
  \centering
  \includegraphics[width=1\textwidth]{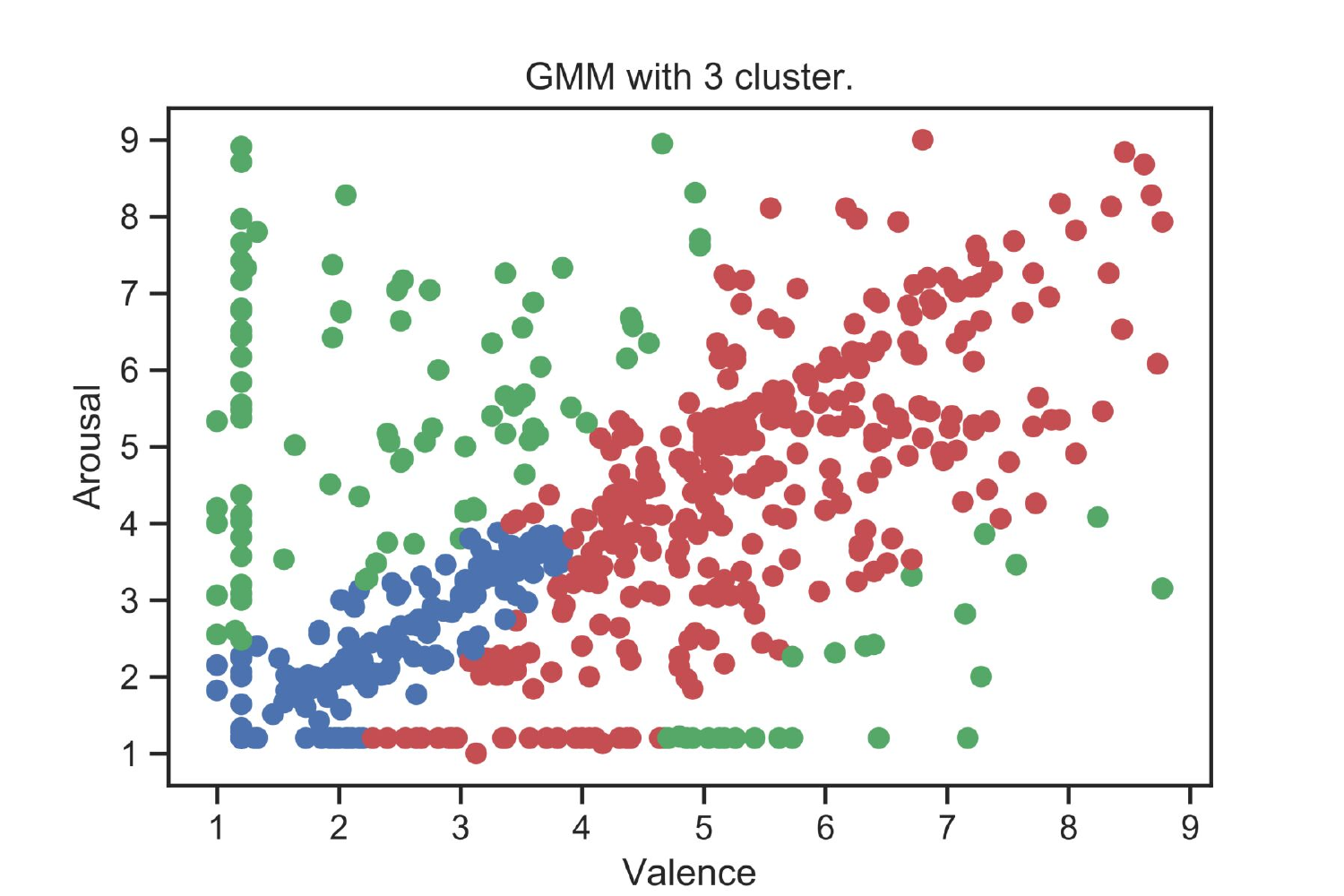}
  \caption{}
  \label{fig:d}
\end{subfigure}
\begin{subfigure}{4cm}
  \centering
  \includegraphics[width=1\textwidth]{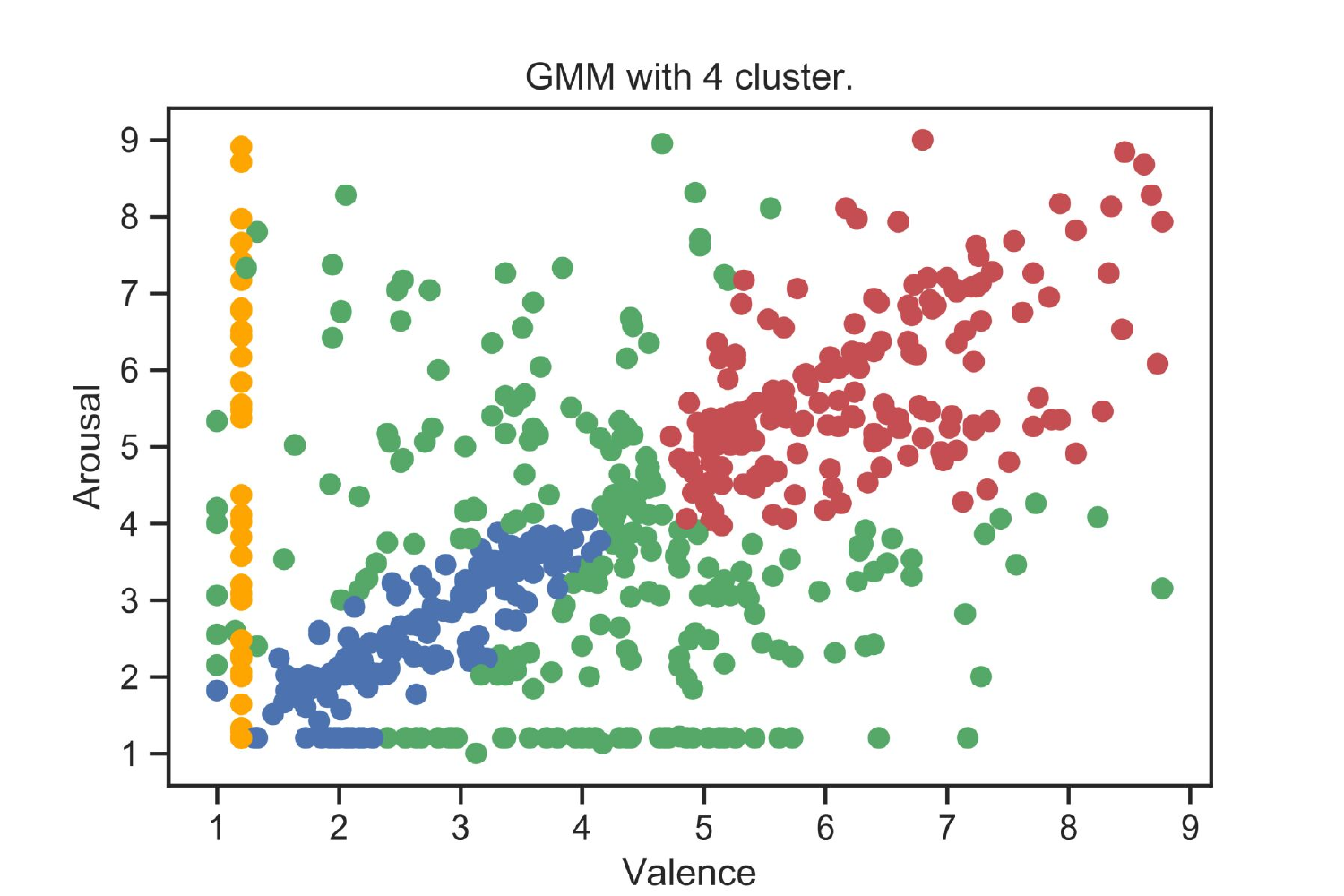}
  \caption{}
  \label{fig:e}
\end{subfigure}
\begin{subfigure}{4cm}
  \centering
  \includegraphics[width=1\textwidth]{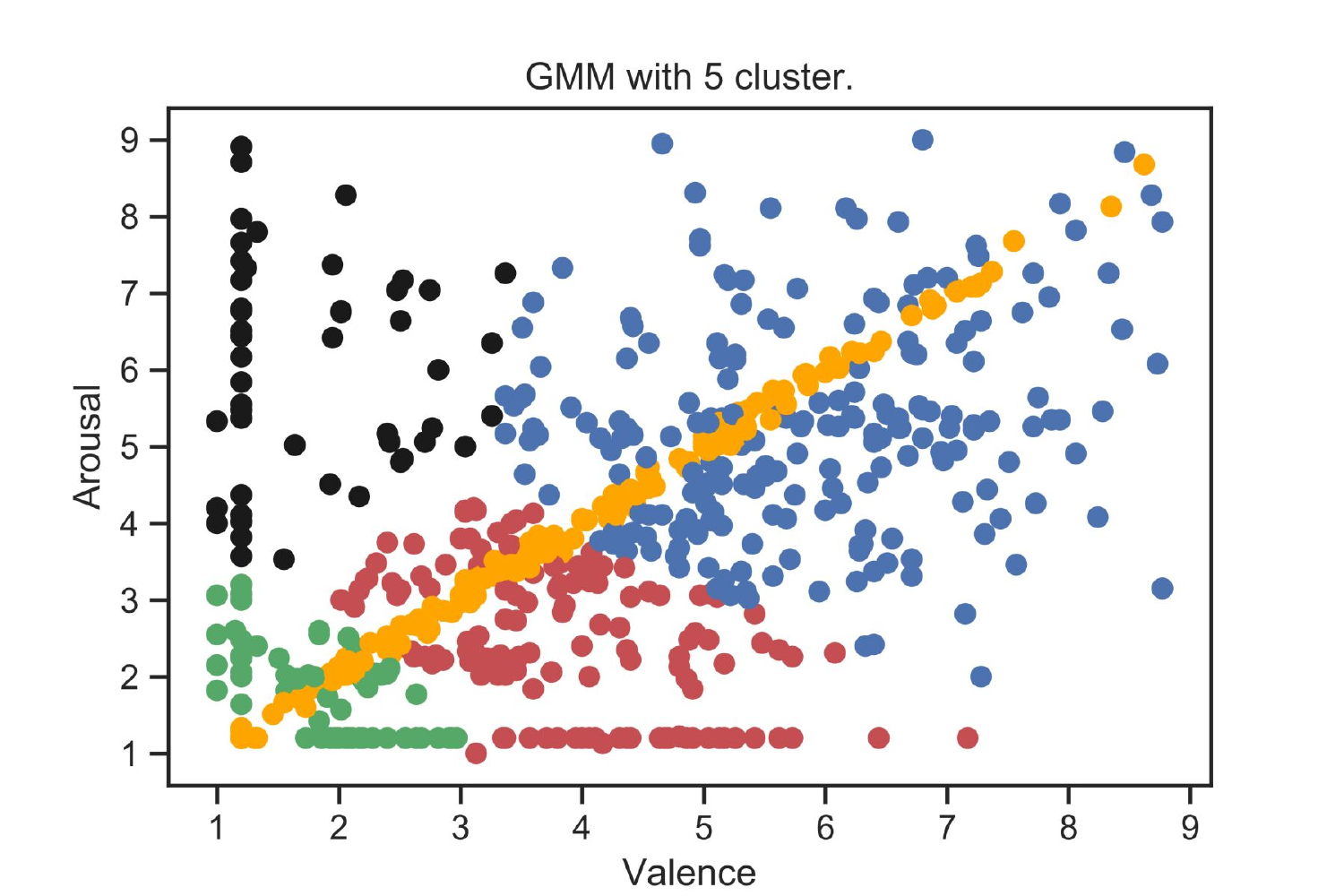}
  \caption{}
  \label{fig:f}
\end{subfigure}
\begin{subfigure}{4cm}
  \centering
  \includegraphics[width=1\textwidth]{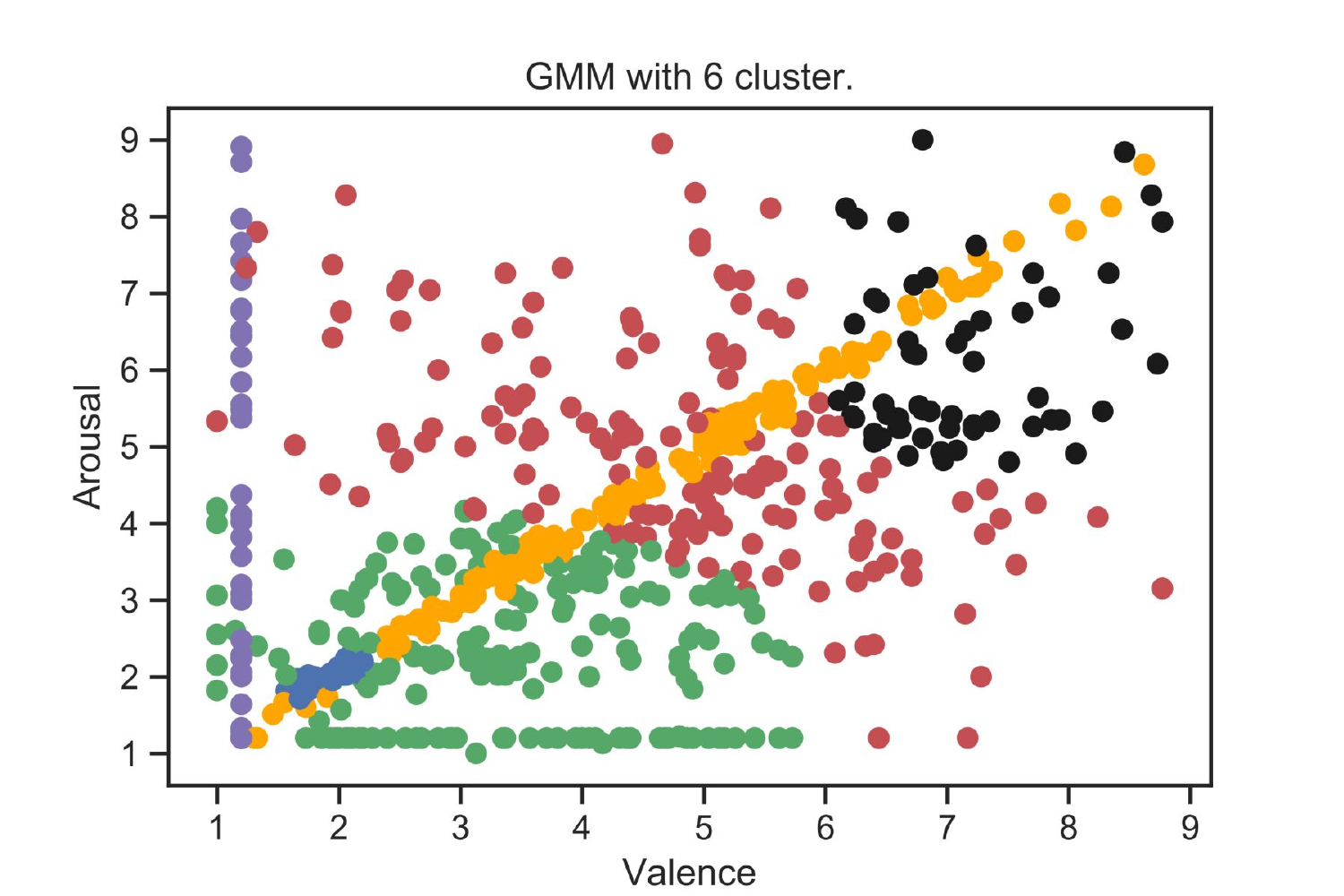}
  \caption{}
  \label{fig:g}
\end{subfigure}
\begin{subfigure}{4cm}
  \centering
  \includegraphics[width=1\textwidth]{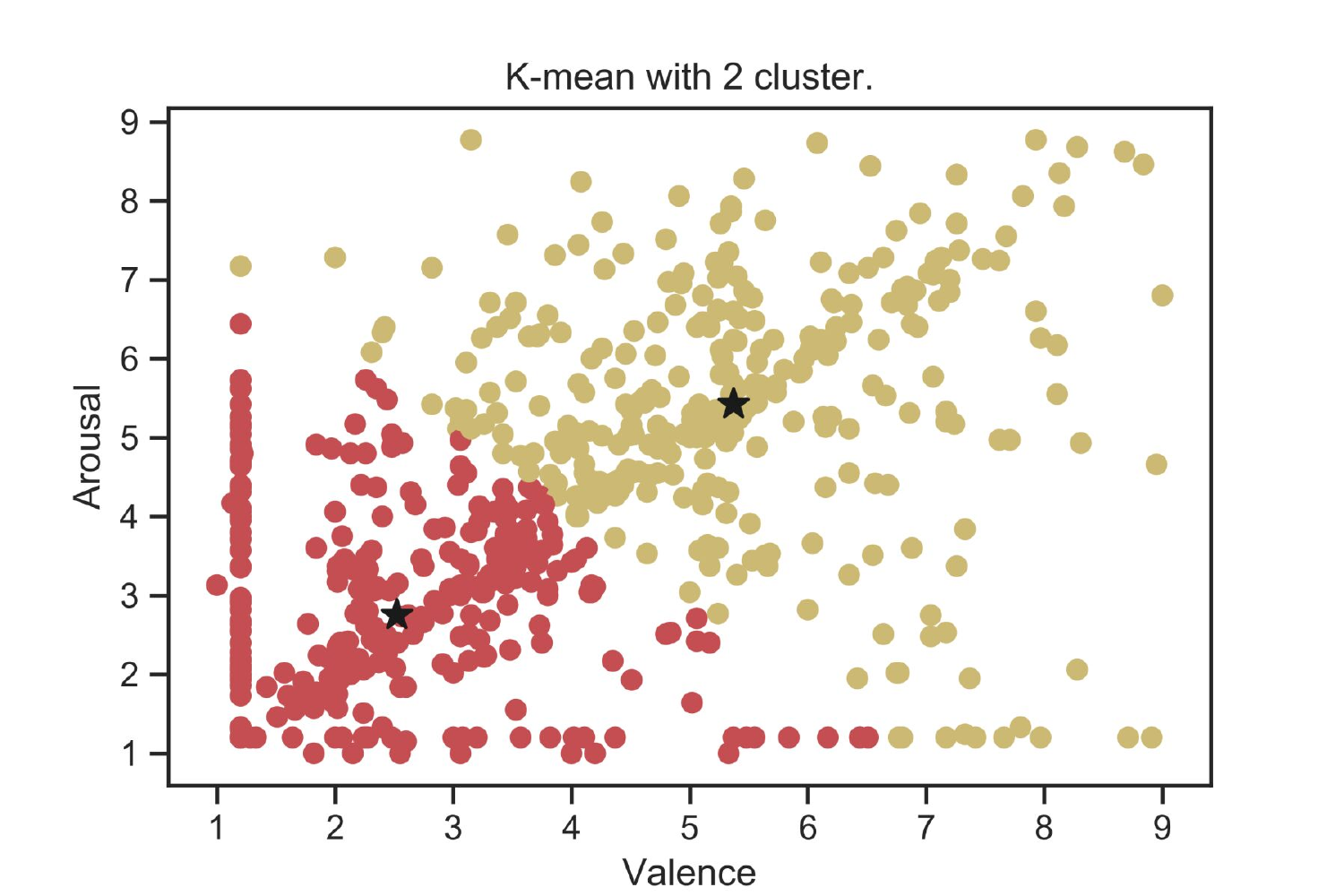}
  \caption{}
  \label{fig:h}
\end{subfigure}
\begin{subfigure}{4cm}
  \centering
  \includegraphics[width=1\textwidth]{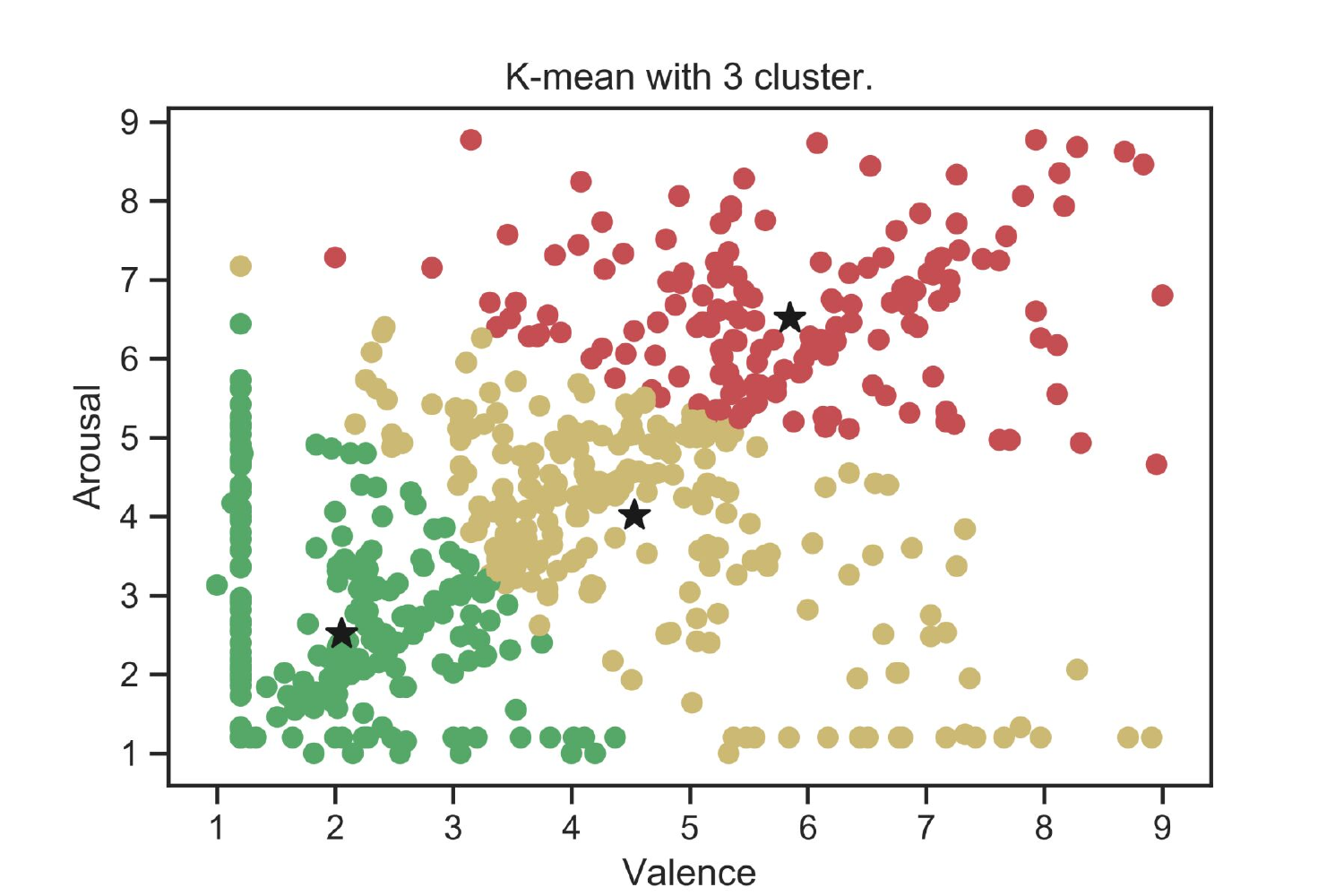}
  \caption{}
  \label{fig:i}
\end{subfigure}
\begin{subfigure}{4cm}
  \centering
  \includegraphics[width=1\textwidth]{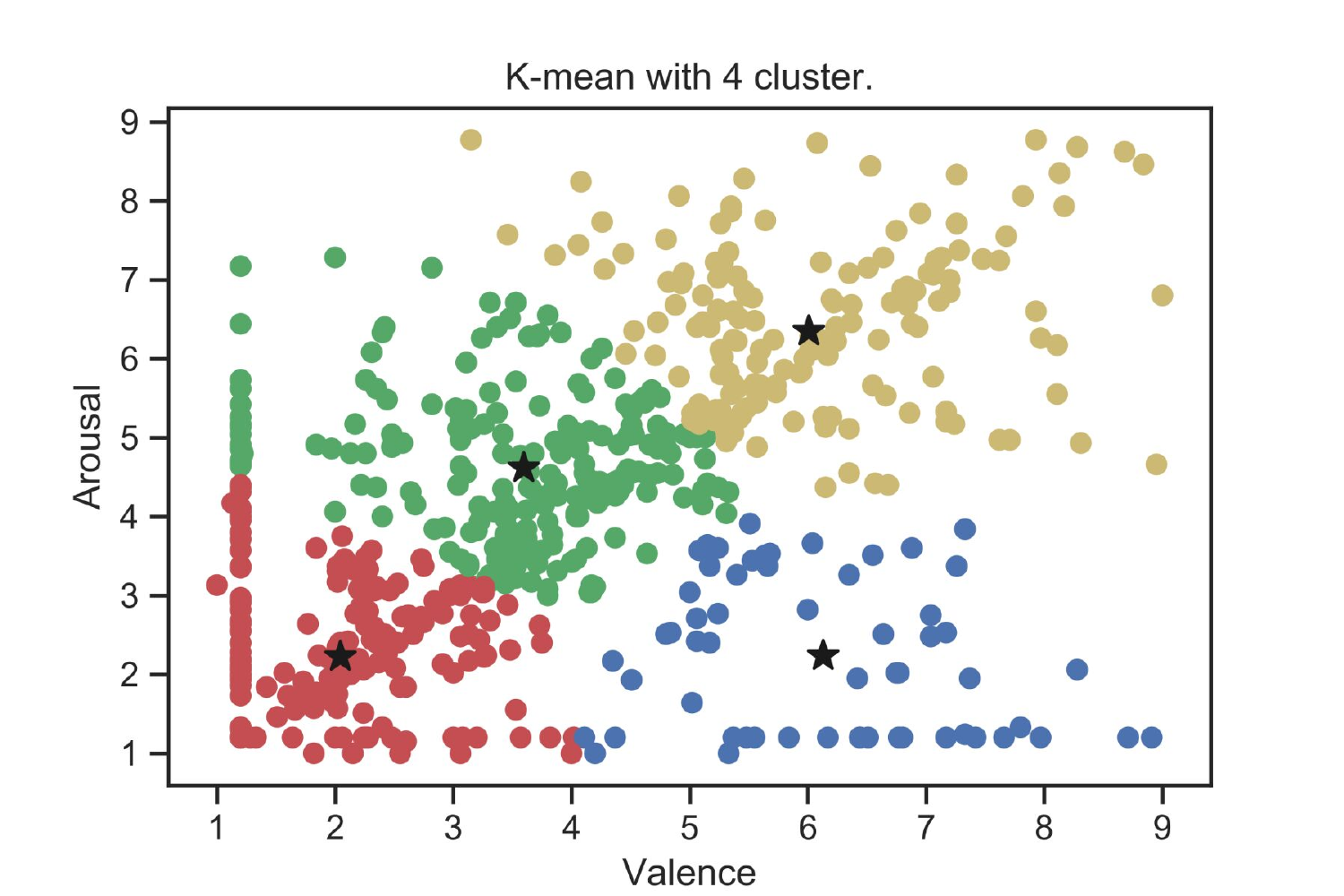}
  \caption{}
  \label{fig:j}
\end{subfigure}
\begin{subfigure}{4cm}
  \centering
  \includegraphics[width=1\textwidth]{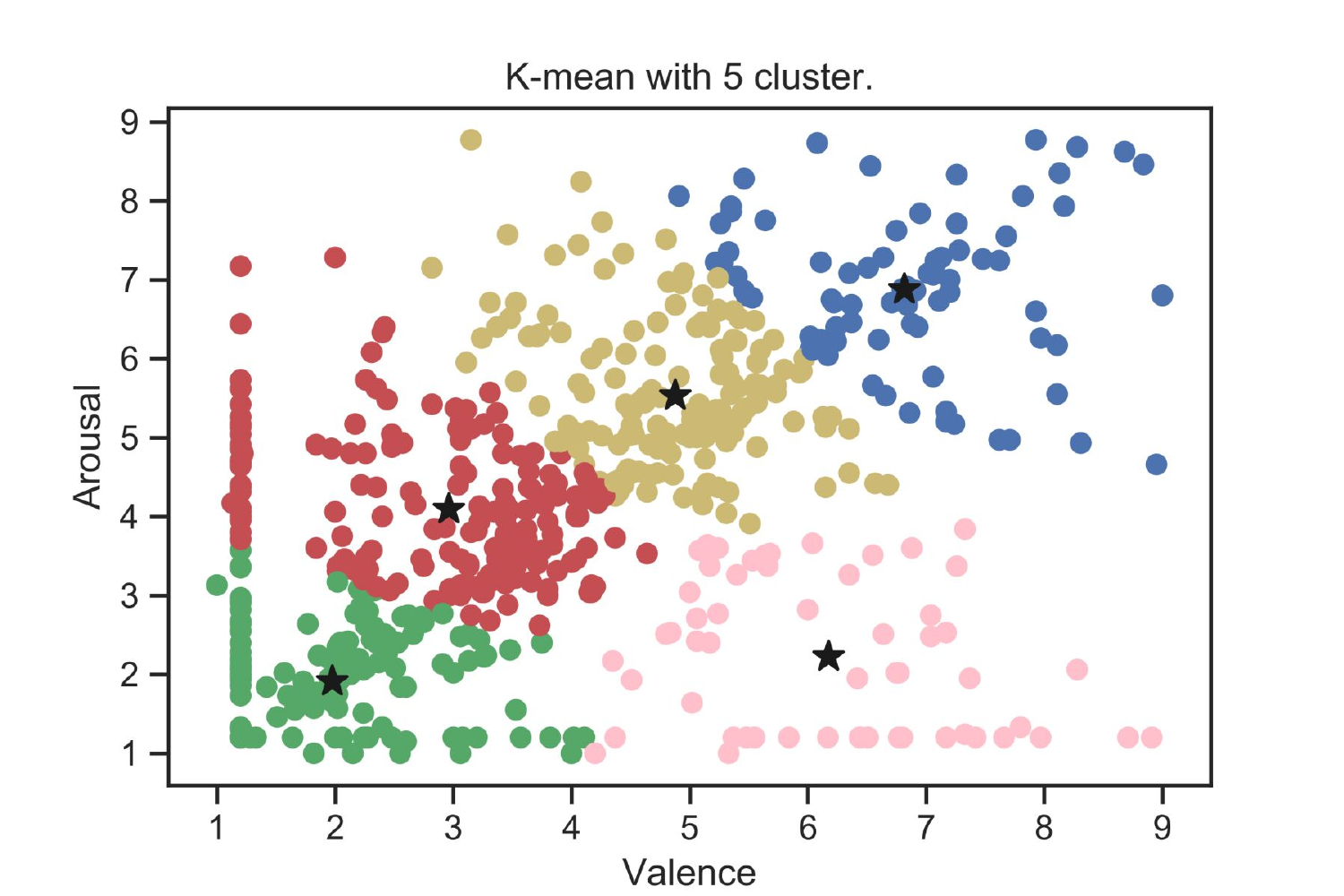}
  \caption{}
  \label{fig:k}
\end{subfigure}
\begin{subfigure}{4cm}
  \centering
  \includegraphics[width=1\textwidth]{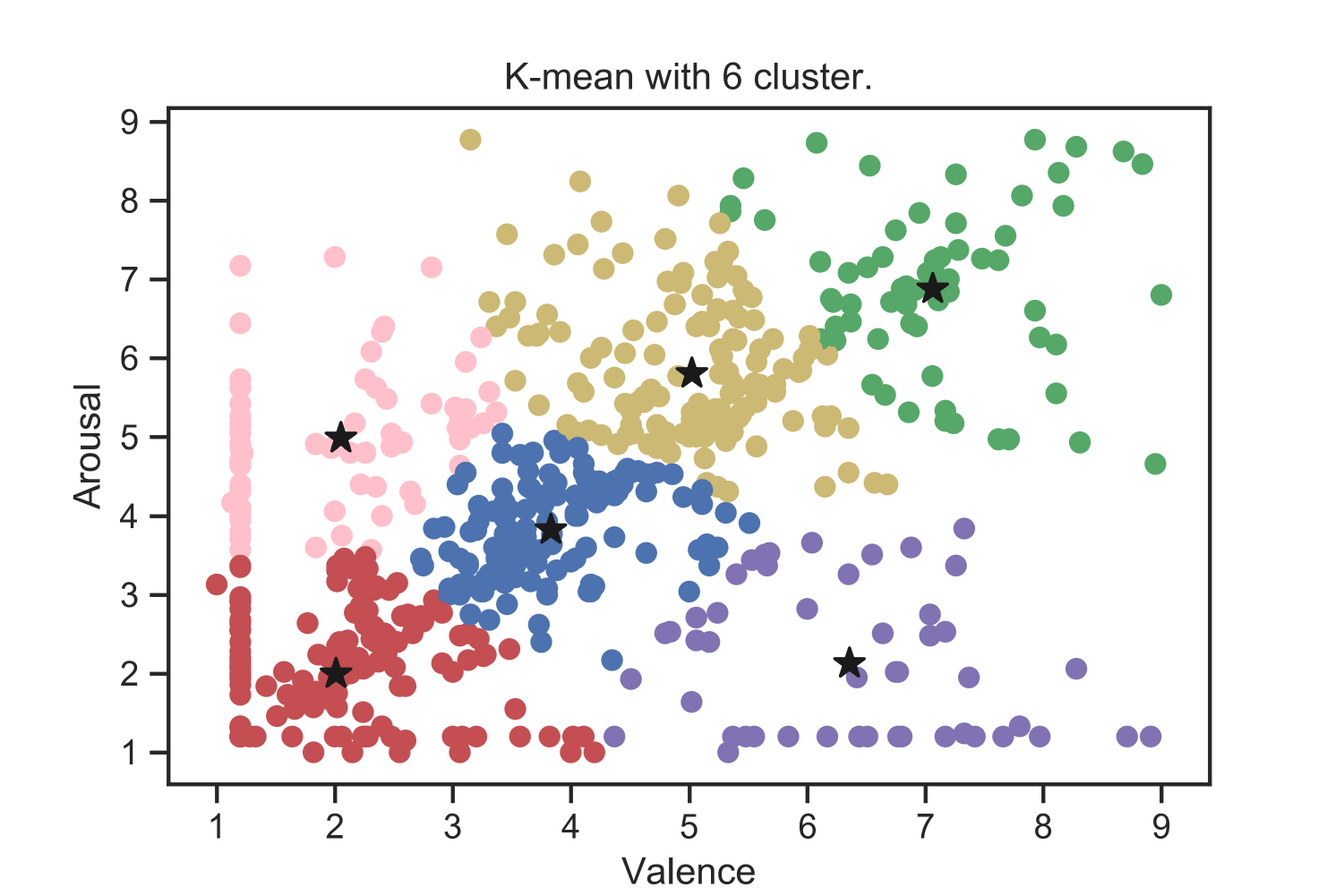}
  \caption{}
  \label{fig:l}
\end{subfigure}
\caption{ (a)-(e) GMM and (f)-(j) K-means plots with increasing number of clusters (K=2 to K=6).}
\label{fig_AnotherMedthod} 
\end{figure}

\begin{table}
\centering
\scriptsize
\begin{threeparttable}
\caption{Comparative table featuring accuracy and $F_{1}$ score for low-high classification, along with results from public repositories that applied identical feature extraction and recognition method in their analysis.} 
\begin{tabular}{l l l p{0.01cm} l c }
\toprule
\toprule
 & \multicolumn{2}{c}{Accuracy}  & &\multicolumn{2}{c}{$F_{1}$ Score} \\
Modality & Valence & Arousal && Valence & Arousal \\
\midrule 
OpenBCI (EEG) & \textbf{0.6667}	& 0.6744 && \textbf{0.6440}	& 0.6470\\
E4 & 0.6383	& 0.6228	&& 0.6200	&0.5912\\
OpenBCI \& E4 & 0.6538	& \textbf{0.7157}	&&0.6428	&\textbf{0.6916}\\
Class Ratio & 1.38:1& 1.66:1&& 1.38:1& 1.66:1\\
\midrule 
OpenBCI (EEG) :K-means  &  0.7080 & 0.6331 && 0.6767 & 0.6233 \\
E4 :K-means  & 0.6357 & 0.6072 && 0.5959 & 0.5761 \\
OpenBCI \& E4 :K-means & \textbf{0.7235} & \textbf{0.6847} && \textbf{0.6932} & \textbf{0.6620} \\
Class Ratio & 1.90:1& 0.68:1&& 1.90:1& 0.68:1\\
\midrule 

DREAMER (EEG)\cite{dreamer} &0.6249&0.6217&&0.5184&0.5767\\
DREAMER (ECG) \cite{dreamer} &0.6237&0.6237&&0.5305&0.5798\\
DREAMER (Fusion) \cite{dreamer} &0.6184&0.6232&&0.5213&0.5750\\
DEAP (EEG) \cite{deap} &0.5760&0.6200&&0.5630&0.5830\\
DEAP (Peripheral) \cite{deap} &0.6270&0.5700&&0.6080&0.5330\\
MAHNOB-HCI (EEG) \cite{mahnob} &0.5700&0.5240&&0.5600&0.4200\\
MAHNOB-HCI (Peripheral) \cite{mahnob} &0.4550&0.4620&&0.3900&0.3800\\
DECAF (MEG) \cite{decaf} &0.5900&0.6200&&0.5500&0.5800\\
DECAF (Peripheral) \cite{decaf} &0.6000&0.5500&&0.5900&0.5400\\
\bottomrule 
\bottomrule
\end{tabular}
\begin{tablenotes}
\small 
\item \textit{Class ratios displayed as [low : high].\\ Optimal values among respective groups shown in boldface}
\end{tablenotes}
\label{table compare}
\end{threeparttable}
\end{table}

\begin{table}
\centering
\scriptsize
\begin{threeparttable}
\centering 
\caption{Accuracy and $F_{1}$ scores associated with channel selection.} 
\begin{tabular}{l c c p{0.1cm} c c } 
\toprule
\toprule
 & \multicolumn{2}{c}{Accuracy}  & &\multicolumn{2}{c}{$F_{1}$ Score}   \\
Channel & Valence & Arousal && Valence & Arousal\\
\midrule 
$FP_1$, $FP_2$, $F_z$ & 0.6331 & 0.5814 && 0.6044 & 0.5590 \\
$FP_1$, $FP_2$, $C_z$ & 0.6770 & 0.6047 && 0.6550 & 0.5934 \\
$FP_1$, $FP_2$, $P_z$ & 0.6409 & 0.5788 && 0.6143 & 0.5565 \\
$FP_1$, $FP_2$, $O_z$ & 0.5917 & 0.6021 && 0.5445 & 0.5681 \\
$T_3$, $T_4$, $F_z$ &  0.6331 & 0.6227 &&  0.5963 & 0.6150 \\
$T_3$, $T_4$, $C_z$ & 0.6486 & 0.6202 && 0.5898 & 0.6135 \\
$T_3$, $T_4$, $P_z$ & 0.6486 & 0.6279 && 0.6119 & 0.6175 \\
$T_3$, $T_4$, $O_z$ & 0.460 &  \textbf{0.6357} && 0.5875 & \textbf{0.6258}\\
$F_z$, $C_z$, $P_z$, $O_z$ &\textbf{0.7184} & 0.5840 && \textbf{0.6741} &  0.5667\\

\bottomrule 
\bottomrule
\end{tabular}
\begin{tablenotes}
\small \item \textit{Optimal values column-wise shown in boldface}
\label{table ch}
\end{tablenotes}
\end{threeparttable}
\end{table}

\begin{table}[h]
\centering
\scriptsize
\begin{threeparttable}
\centering
\caption{Accuracy and $F_{1}$ scores from different EEG frequency bands.} 
\begin{tabular}{l c c p{0.1cm} c c } 
\toprule
\toprule
 & \multicolumn{2}{c}{Accuracy}  & &\multicolumn{2}{c}{$F_{1}$ Score}   \\
Frequency Band & Valence & Arousal && Valence & Arousal\\
\midrule 

$\theta$ & 0.6718 & 0.6770 && 0.6363 & 0.6518 \\
$\alpha$  & \textbf{0.7003} & 0.6434 && \textbf{0.6640} & 0.6123 \\
$\beta$ & 0.6925 & \textbf{0.6796} && 0.6557 & \textbf{0.6590} \\
$\gamma$ & 0.6719 & 0.6667 && 0.6286 & 0.6394 \\

\bottomrule 
\bottomrule
\end{tabular}
\begin{tablenotes}
\small \item \textit{Optimal values column-wise shown in boldface}
\label{table freq}
\end{tablenotes}
\end{threeparttable}
\end{table}

\subsection{Experiment I: Affective Video Selection}
To start off, we analyzed the participants' self ratings on Happiness (H), Fear (F), and Excitement (E). In \autoref{fig4} (a), all data points are scattered across a three-dimensional space (H, F, E). We adopted K-means clustering and GMM methods and employed Davies-Bouldin index (DB-index) and Elbow method to select the optimal K parameter. The outcomes are shown in \autoref{tableDBindex1}; the minimal or best DB-index achieved corresponded to K=3 in K-means clustering. Moreover, in \autoref{figElbow1}, the Elbow method further confirmed the optimal number of clusters for K-Means to be three. In \autoref{fig4}, all data points were assigned a certain color according to their designated cluster, as calculated by the K-means method (see (b)). The points corresponding to clips that were not allocated in the same cluster as their majority class (i.e., mode) were filtered out. Following which, the scatter points in each cluster were obtained as illustrated in \autoref{fig4} (c). Lastly, only 20 points in each cluster (60 clips in total) were retained to work out the nearest distances to the centroids (see \autoref{fig4} (d)). We hypothesized that the clips (\autoref{tableclip}) chosen by the assistance of unsupervised learning would prove to be effective tools for emotion induction exercise.

\subsection{Experiment II: Emotion Recognition Using OpenBCI}
As outlined in Methodology, our investigation was divided into three subtasks. The following paragraph provides details on the experimental results of these subtasks.

\begin{itemize}
\item \textbf{Low-High: Valence/Arousal with Threshold}

For this analysis, ground truth labels were generated based on self-emotional assessment scores (V and A) and the threshold empirically set at 4.5. Scores higher than the threshold were assigned the high-level label, and vice versa for any scores lower than the threshold. The input features extracted from EEG, E4, and Fusion (EEG and E4) were used to train and test the model. \autoref{table compare} presents the mean accuracy and mean nine-fold $F_{1}$ score. The condition with EEG as input features reached 66.67 \% of mean accuracy and 0.6640 of mean $F_{1}$ score for A. For V, the condition with Fusion of EEG and E4 as input features reached 71.57 \% of mean accuracy and 0.6916 of mean $F_{1}$ score. Furthermore, we performed repeated measure ANOVA, with Greenhouse-Geisser correction for statistical analysis. It was found that the mean accuracy of low-high A classification was significantly different among the application of EEG, E4, and Fusion data (F(1.994, 15.953)=15.791), p $<$ 0.05,  df(factor, error)=1.994,15.953). Further, the Bonferroni post hoc test revealed that Fusion (71.574 $\pm$ 1.578\%) was significantly better than using either EEG only (67.442 $\pm$ 1.644\%) or E4 only (62.866 $\pm$ 0.253\%).

\item \textbf{Low-High: Valence/Arousal with Clustering}

As shown in \autoref{tableclip}, V and A scores appeared to be widely distributed given their considerable standard deviation values. It could be the case that larger population is by default associated with higher standard deviation. Hence, the fixed threshold strategy might not be suitable in this scenario. To address this potential issue, we tested two clustering methods, K-Means and GMM. As shown in \autoref{tableDBindex2}, K-means clustering with 4 clusters was likely the most suitable method as indicated by its minimal DB index. Moreover, as shown in \autoref{fig_AnotherMedthod}, comparing the clustering results between these methods and different numbers of clusters, our chosen solution (h) was better than the others. Therefore, it was adopted for the application of labeling low-high levels of V and A. We empirically labeled according to the VA model \cite{russell1980}, with the Blue and Red group for low-A (LA), and the Green and Yellow group for high-A (HA). In terms of valence, the Red and Green group was selected for low-V (LV) and the Blue and Yellow group for high-V (HV).
After that, binary classification was carried out, the resulting mean accuracy and mean nine-fold $F_{1}$ scores are laid out in \autoref{table compare}. The condition with the Fusion of EEG and E4 as the input features reached the mean accuracy of 72.35\% and 68.47\% for V and A, respectively. In terms of $F_{1}$ score, Fusion features provided the best results for V and A (0.6932 and 0.6620, respectively). We also performed repeated measures ANOVA with Greenhouse-Geisser correction. It appeared that the mean accuracy of low-high A classification was significantly different between using EEG, E4, and Fusion (F(1.574, 12.591)=10.057), p $<$ 0.05. Also, the Bonferroni post hoc test revealed that using Fusion (72.350 $\pm$ 1.191\%) was significantly better than using only E4 (63.567 $\pm$ 1.450\%). \autoref{table compare} reports the V and A classification results from DREAMER\cite{dreamer}, DEAP\cite{deap}, MAHNOB-HCI\cite{mahnob}, and DECAF\cite{decaf} datasets all using identical feature extraction and classification method. The observed performance of our data seemed to be lending support to the capability of the consumer grade brain sensor, OpenBCI, as a versatile tool in emotion recognition studies.

\item \textbf{EEG Electrode Channels and Frequency Bands}

The results from channel selection are presented in \autoref{table ch} with the mean accuracy and mean $F_{1}$ score from nine folds. In the low-high classification task for V, the condition with EEG channels from $F_z$, $C_z$, $P_z$, $O_z$ as the input features achieved a mean accuracy of 71.84\% and a mean $F_{1}$ score of 0.6741. However, for A, the mean accuracy reached 63.57\% and the mean $F_{1}$ score reached 0.6258 using $T_3$, $T_4$, $O_z$. Finally, the results obtained from varying EEG frequency bands, as shown in \autoref{table freq}, indicate the mean accuracy and mean $F_{1}$ score from nine folds. The classification accuracy for V reached 70.03\%  by using EEG features from $\alpha$ band and A reached 67.96\% by using EEG features from $\beta$ band. In terms of $F_{1}$ score, the accuracy results were similar. The features from $\alpha$ bands provided the best results in V and the features from  $\beta$ band provided the best results in A (0.6640 and 0.6590, respectively). We then carried out repeated measures ANOVA, however, there appeared to be no significant difference between the selection results of frequency and channel.

\end{itemize}

\section{Discussion}
The practice of evoking emotions on command is commonplace in emotion research and lately has gained a foothold in extended applications such as classification of affective states from correlated EEG and peripheral physiological signals. A number of previous datasets have been generated using audio-visual stimuli, more specifically video clips, to invoke desired emotions in individuals while at the same time record their EEG and bodily responses. In the present study, we propose a novel method for affective stimulus selection, involving video ratings by 200 participants aged 15--22 and K-means clustering method in video ranking. A complete list of movie titles used in our study is provided in \autoref{tableclip}. These videos can readily be found on YouTube and online websites, should there be a need for further use.

The emotion recognition task in this study resorted to the measurement of valence and arousal experienced by the viewers during video presentation. In most studies, ground truth labeling relies on an empirically set threshold for high-low partitioning \cite{deap,mahnob,dreamer,decaf}. However, the list of selected titles in \autoref{tableclip} shows large standard deviations for self-assessment scores. In a large population, simple thresholding might not work well since different people are likely to experience varying levels of low or high valence and arousal. Thus, a simple mathematical based algorithm, K-means clustering, was proposed for labeling the ground truth. As reported in \autoref{table compare}, the proposed method for ground truth labeling may lend support to attaining high classification performance, especially in valence classification. Moreover, the valence and arousal classification results among using EEG, peripheral signals, and fusion were notably consistent compared to the conventional ground truth labeling method. Hence, labeling by K-means clustering might be more suitable in future emotional studies with a large number of participants. Moreover, the outcome of emotion recognition from OpenBCI was comparable to state-of-the-art works featuring high-end or expensive EEGs (both in accuracy and $F_{1}$ score) as can be seen in \autoref{table compare}, including one involving MEG, a functional brain mapping method (considerably more expensive than EEG).

\autoref{table ch} and \autoref{table freq} display the study results of EEG factors -- i.e., electrode channel selection for future development of user-friendly devices in emotion recognition, and frequency selection to identify more effective EEG frequency bands as classification features. Referring to the tables, $T_3$, $T_4$, $O_z$ achieved the best results for arousal classification both in terms of accuracy and $F_{1}$ score, while the middle line ($F_z$, $C_z$, $P_z$, $O_z$) was promising for further improvement of an emotion recognition algorithm in valence for low-high classification. Taking both results together, $T_3$, $T_4$, $O_z$, $F_z$, $C_z$, $P_z$ with all EEG frequency bands as input features offer the best path for developing practical, usable consumer grade devices for mental state recognition, especially one concerning valence and arousal. True to any scientific investigation, there are limitations inherent to this study, one of which comes down to ICA implementation. We were able to pinpoint the axes with EMG traits, however, decided not to remove any of them as this may lead to a potential loss of important information.

To our knowledge, this study is the first to carry out an evaluation of OpenBCI applicability in the domain of emotion recognition. In comparison to medical grade EEG amplifiers with a greater number of electrodes and sampling frequencies, OpenBCI demonstrably could hold its own. A consumer grade, open-source device has a potential to be a real game changer for programmers or researchers on the quest for better emotion recognition tools. The device may facilitate further progress toward online applications since it is inexpensive and possibly affordable even to those with more limited purchasing power. In the same vein, emotion recognition using peripheral physiological data from a real-time bracelet sensor or E4 remains a challenge. E4 is an easy-to-use wearable device resembling a watch which may be useful as a long-term affective state or mental health monitoring system. In fact, a research team recently proposed the algorithm K-nearest neighbors, based on dynamic time warping, to process physiological signals and perform affective computing on E4-derived datasets\cite{toward}. We might also be able to adapt this work for OpenBCI. Besides, we have two ongoing projects on deep learning and EEG in which the data from this study could be incorporated for further investigation \cite{affective,universal}.

\section{Conclusion}
We present evidence in support of the applicability of OpenBCI, an open-source, consumer grade EEG device for emotion recognition research. A brief summary of our effort to collect this evidence is as follows. EEG and peripheral physiological signals were collected from 43 participants as they were being shown a sequence of short movie trailers expected to invoke distinctive emotions. The participants were prompted to score their experienced valence and arousal for each video. Subsequently, we applied K-means clustering algorithm on these valence and arousal ratings in order to establish ground truth labels for low and high cluster, dissimilar to reported studies that performed labeling by empirical thresholding. We found that our prediction outcome appeared in a similar performance range to those derived from state-of-the-art datasets that were acquired by medical grade EEG systems. The ultimate goal of this study is to inform and inspire researchers and engineers alike on the practicality of OpenBCI as a recording device in the development of online, emotional EEG-related applications. Our experimental data are available to academic peers and researchers upon request.

\section*{Acknowledgment}
The authors would like to thank Sombat Ketrat for his support and suggestions on server installation for data storage and computing. Our thank also extends to Irawadee Thawornbut for her assistance in the setting up of the preliminary study.

\ifCLASSOPTIONcaptionsoff
  \newpage
\fi

\bibliographystyle{IEEEtran}
\bibliography{ref} 
\end{document}